%% file: main.tex
\documentclass[a4paper,11pt]{article}
\usepackage{jinstpub} 

\usepackage{orcidlink}
\usepackage{graphicx}
\usepackage{pdflscape}

\usepackage{lineno}

\title{\boldmath
Studies of Hadronic Showers in SND@LHC}

\collaboration{The SND@LHC Collaboration}

\emailAdd{marco$\_$dallavalle@cern.ch}

\abstract{
The SND@LHC experiment was built for observing neutrinos arising from LHC pp collisions. The detector consists of two sections: a target instrumented with SciFi modules and a hadronic calorimeter/muon detector. Energetic $\nu$N collisions in the target produce hadronic showers. 
Reconstruction of the shower total energy requires an estimate of the fractions deposited in both the target and the calorimeter. 
In order to calibrate the SND@LHC response,  
a replica of the detector was exposed to hadron beams with 100 to 300 GeV in the CERN SPS H8 test beam line in Summer 2023.
This report describes the methods developed to tag the presence of a shower, to locate the shower origin in the target, and to combine the target SciFi and the calorimeter signals so to measure the shower total energy.
}

\begin{document}
\include{Authorlist}

\maketitle
\flushbottom


\section{Introduction}
\label{sec:intro}
The SND@LHC experiment detects high energy neutrinos emerging from LHC collisions at very small angles~\cite{Albanese:031802}.
For all neutrino flavors, energetic $\nu$N collisions produce hadronic showers. 
The hadronic shower energy is one of the observables, together with shower direction and charged lepton track parameters, needed for estimating the incoming neutrino energy in Charged Current 
$\nu$N interactions.

This report describes the energy calibration for the SND@LHC detector that was performed with hadron test beams at the CERN SPS in 2023. The detector consists of a massive target section complemented with a calorimeter section (HCAL) and a muon detector.
The reconstruction of the shower total energy requires to estimate the fraction lost in both the target and the calorimeter.   
The energy sharing also depends 
on the position of the neutrino interaction vertex along the target depth. 

After a description of the test beam setup, the report develops in four steps: (i)~ the construction of an algorithm to tag the presence of a particle shower and to locate the origin along the target, 
(ii)~the study of the correlation between energy losses in target and HCAL, and their recombination for reconstructing the shower total energy,
(iii)~the analysis of the range of applicability of the shower energy reconstruction method,
and, finally, (iv)~the comparison with the performance expected from Monte-Carlo simulations.


\section{Test Beam Setup}
\label{sec:detector}

The goal of the SPS beam test was to calibrate the hadronic calorimeter (HCAL) response of the SND@LHC experiment. 
Therefore, detectors, frontend (FE) and data acquisition (DAQ) electronics were built to duplicate the hardware installed in 2022 in the LHC TI18 tunnel in every detail.

\begin{figure}[tb]
\centering
\includegraphics[width=1.0\textwidth]{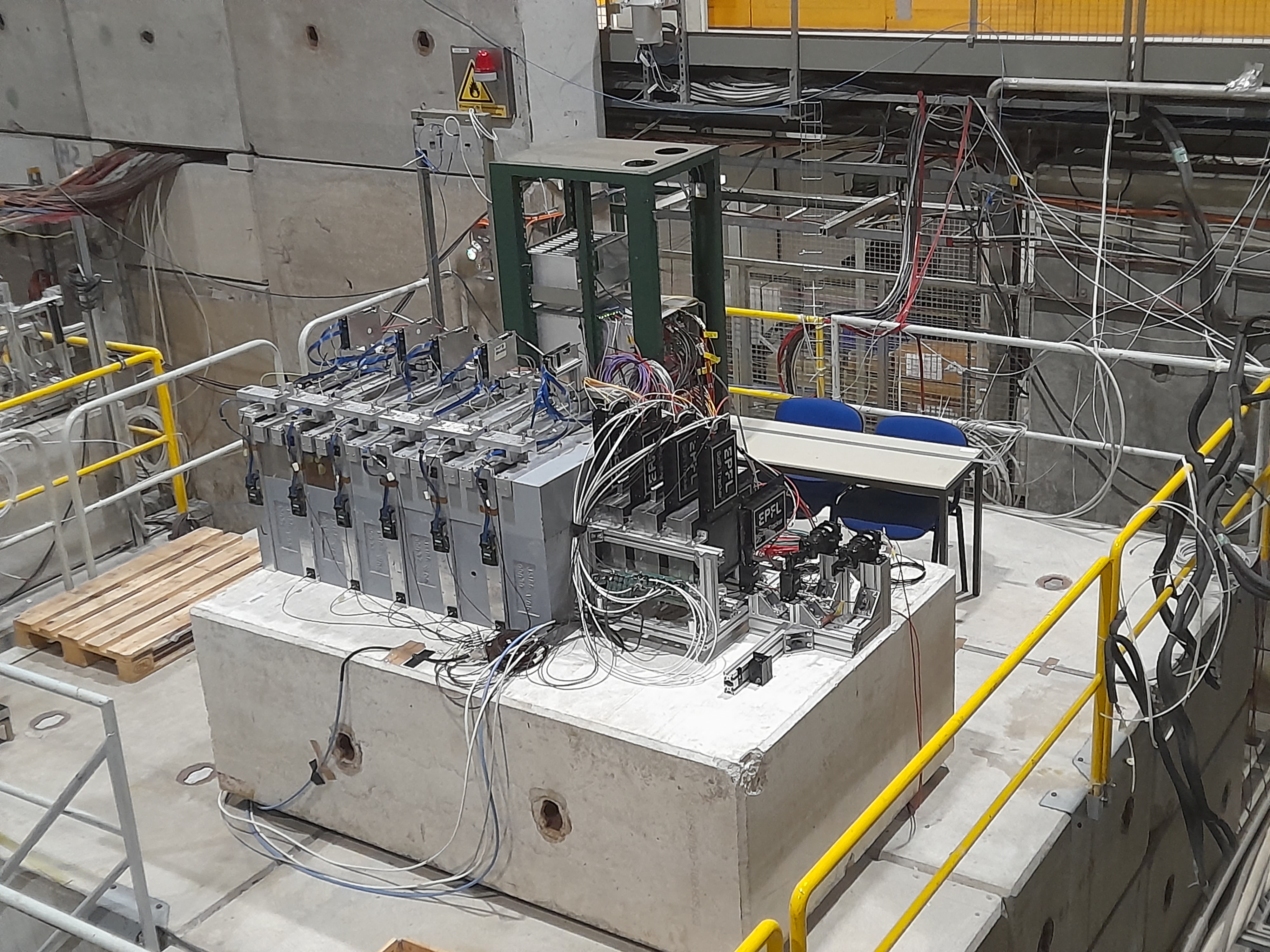}
\caption{
Detector setup in the SPS H8 test beam line with the beam coming from the right.
From right to left: (i) small-size Beam Counter scintillators, (ii) three target walls interleaved with four SciFi X-Y detectors, and (iii) the hadronic calorimeter (HCAL), consisting of planes of scintillating bars interspaced with (grey) iron walls. }
\label{fig:DetectorOnBeam}
\end{figure}

In Summer 2023 the detector was exposed to hadron beams with energies in the  range of 100 to 300 GeV in the SPS H8 test beam line
(figure~\ref{fig:DetectorOnBeam} )

\subsection{Hadronic Calorimeter (HCAL)}
\label{subsec:HadCalo}

\begin{figure}[tb]
\centering
\includegraphics[width=1.0\textwidth]{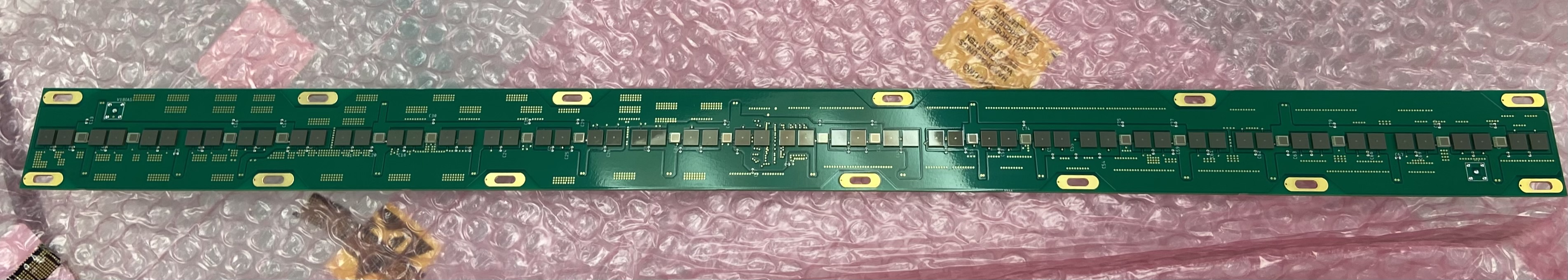}
\caption{ PCB hosting the SiPMs that read-out the bars on one side of a US station.}
\label{fig:USPCB}
\end{figure}

Five Upstream (US) stations~\cite{Acampora2024}
of large scintillating bars constituted the sensitive part in the HCAL, and were interleaved with $20\ \rm cm$ thick iron walls. 
The US stations consisted each of ten bars of plastic ELJEN EJ 200, $60\ \rm mm$ wide, $830\ \rm mm$ long, $10\ \rm mm$ thick.
The bars were tightly wrapped in $20\ \rm \mu m$ thick aluminized mylar foils, and framed side by side between two $2\ \rm mm$ thick aluminum sheets, to constitute a station of $830\ \rm mm$ width and $600\ \rm mm$ height.
Each bar  was read-out, at both left and right edge, with 6 "large" silicon photomultipliers (SiPM) (Hamamatsu Photonics MPPC S14160-6050HS, $6 \times 6~\mathrm{mm}^2$) 
and 2 "small" SiPMs (Hamamatsu Photonics MPPC S14160-3050H, $3 \times 3~\mathrm{mm}^2$).
On both sides of  a US station, the 60 large and 10 small SiPMs were hosted on a single PCB
(figure~\ref{fig:USPCB}).
and the PCBs were carefully aligned, so that each bar edge was uniquely read out by 6 large and 2 small SiPMs on each side.

The optical coupling between SiPM and scintillating bars was implemented on one side with gel pads and on the other without, similarly to what was done for the US stations of the SND@LHC detector. 
First, at one edge, bars were pushed in close contact with the SiPMs with no gel, 
then at the opposite end 
a silicon gel pad - made from WACKER SilGel(R) 612 A/B - was used to make efficient contact with the SiPMs. This procedure compensated for small differences between bar lengths. 
Subsequent US stations in HCAL were arranged so that the gel side was alternate left or right.
At both sides of a US station, the SiPMs signals were routed to the center of the PCB, and from there to the FE PCB, mounted on the US station and based on two TOFPET ASICs~\cite{Bugalho_2019} for the readout $30$ SiPMs.
A Downstream (DS) station,  with two planes of  60 thin bars of $10\ \rm mm$ width, was also built and installed downstream the HCAL. In the first DS plane bars were arranged horizontally, in the second vertically.

\begin{figure}[tb]
\centering
\includegraphics[width=1.0\textwidth]{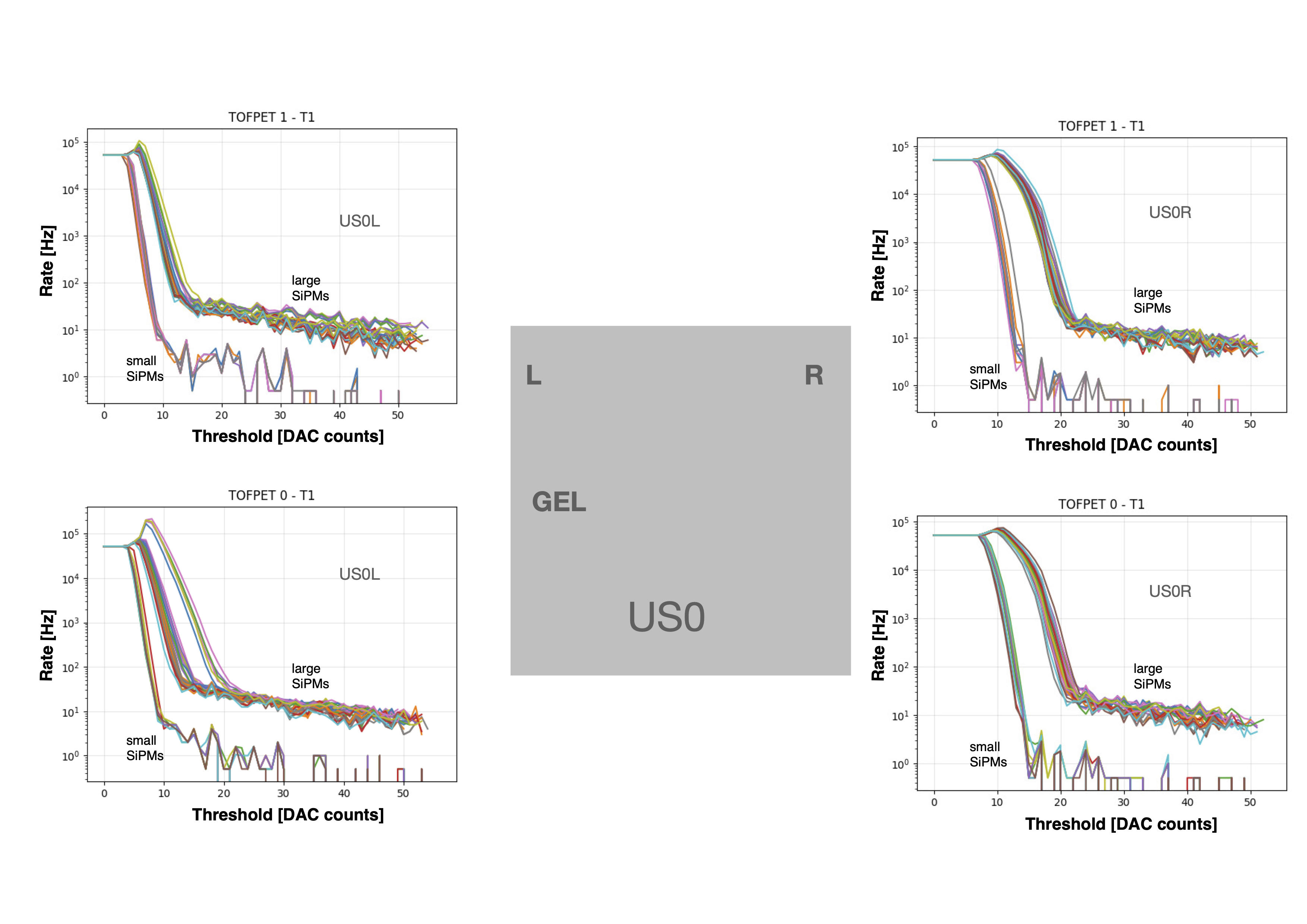}
\caption{ Hit frequency as function of the signal thresholds (T1) in both TOFPETs (0~and~1) of both PCBs (Left and Right) in US0, the most upstream HCAL station in the test beam. On the Left side of US0 a silicon gel pad  was used to make contact between scintillating bars and the SiPMs.
Each plot maps 30 "large" SiPMs and 10 "small" SiPMs.
A few slightly noisier channels show up in the bottom left plot. 
The T1 threshold is set at 30 DAC counts.}
\label{fig:USDCR}
\end{figure}

The complete US0,1,2,3,4 and DS stations were submitted to thorough electrical and light-tightness checks.
The performance was excellent, 
comparable to the detectors installed 
in the SND@LHC.
Of the 800 US SiPMs in 10 PCBs, $99.3\ \%$ were functional: two channels were silent  
and four, being too noisy, were masked.
These six channels were randomly distributed over four PCBs, which where then used to equip the two most downstream stations in the HCAL.
Figure~\ref{fig:USDCR} shows the dark count rate 
as function of signal thresholds for all channels in both TOFPETs of both PCBs in the US0 station. 

Four DAQ boards, spare of the SND@LHC detector~\cite{Acampora2024}, were used to acquire data from the ten FE PCBs of the five US stations.
A test stand with the stations lying horizontally on top of each other was built to acquire cosmic ray data and tune the 
Read-Out synchronization
of the DAQ boards before installation in the SPS H8 beam area.

\subsection{Event Timing}
\label{subsec:hitTiming}

The readout system is asynchronous and triggerless, as in SND@LHC.
Every time any SiPM fires 
(a hit), 
a new hit is recorded, and
this happens 
for each individual DAQ board independently. 
The DAQ system~\cite{EttoresPaper}
of the SND@LHC experiment runs using an internal clock with a frequency of $160.316$~MHz, which translates to clock cycles of approximately $6.25$~ns. 
The DAQ event builder software merges hits in time over all DAQ boards into events.
This grouping is done by looking at the earliest hit time stamp, and aggregating any other hits whose initial timestamp falls 
within 
a predetermined amount of time, with the preset for the LHC detector being four clock cycles. 
This leads to events lasting approximately $25$~ns.
In the case of the test beam setup, 
the event window was increased to sixteen clock cycles, which leads to approximately $100$~ns. 

\subsection{Target Configurations and Data Taking}
\label{subsec:data}

The data taking was planned with the goal to study the detector response as function both of the hadron energy and of the interaction depth in the target.

\begin{figure}[tb]
\centering
\includegraphics[width=1.0\textwidth]{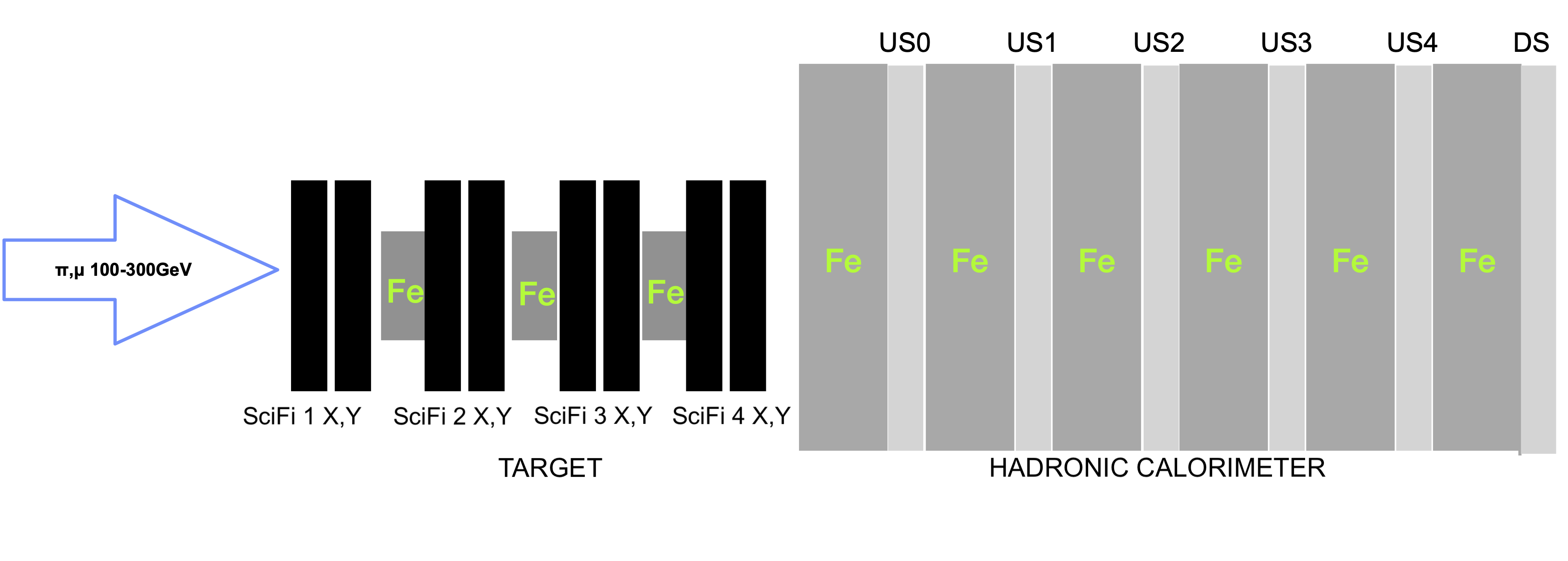}
\caption{ Sketch of the detector on the beam line. From left to right: the Target (three 10~cm thick iron walls interleaved with four SciFi X and Y planes) and the hadronic calorimeter (HCAL) (five US planes of scintillating bars interspaced with 20~cm thick iron walls, and  a DS plane of thin scintillating bars).
The pion interaction length in iron is $20.4\ \rm cm$.}
\label{fig:detector_sketch}
\end{figure}

Figure~\ref{fig:detector_sketch} shows a graphical representation of the detector on beam.
Upstream of HCAL, three $10\ \rm cm$ thick iron walls constituted the target, equipped with four $13 \times 13\ \rm cm^2$ SciFi stations - smaller area replica of the SciFi detectors used in 
SND@LHC~\cite{Acampora2024}
each with a horizontal and a vertical fibers plane, for measuring Y and X coordinates respectively.
Although the target had a reduced size with respect to the SND@LHC, this setup allowed for studying all the features of hadronic showers development, if the beam was centred in the SciFi acceptance.

Data were collected with hadron beams of five energies  and in three configurations of the target, as summarized in table~\ref{tab:TBData}. 
\begin{table} [ht]
\centering\caption{ Number of events collected (in multiples of $10^6\ \rm events$), and beam energies and target configurations used.}
\smallskip
\begin{tabular} { l c c c c c} 
\hline
 & $\pi^+$ & $\pi^+$  & $\pi^+$ 
& $\pi^-$ & $\pi^-$  \\
 & 100 GeV  & 140 GeV   & 180 GeV  
& 240 GeV  & 300 GeV   \\
\hline
3 target walls & 51 &50 &83 &98 &121 \\ 
2 target walls  &15 &15 &30 &61 &61\\ 
1 target walls  &15 &21 &27 &64 &50\\ 
\hline
\end{tabular}
\label{tab:TBData}
\end{table}
Beams of positive hadrons (protons and pions, called $\pi^+$
beam hereafter) at 100, 140, 180 GeV, and of negative pions at 240, 300 GeV were setup at the SPS H8 beam line. 
The beam energy spread was optimised to be narrower than $2\ \%$.
The highest energies were available only with $\pi^-\ \rm beams$, so that some hours were to be allocated in the data taking plan for switching the magnet polarities along the beam line.
Beam intensities were tuned to about $5\times10^3\ \rm particles/spill$, because with higher fluxes, given the large iron mass of HCAL, absorbers would need to be installed around the detector for dumping the radiation background produced by hadronic interactions.

\begin{figure}[tb]
\centering
\includegraphics[width=0.65\textwidth]{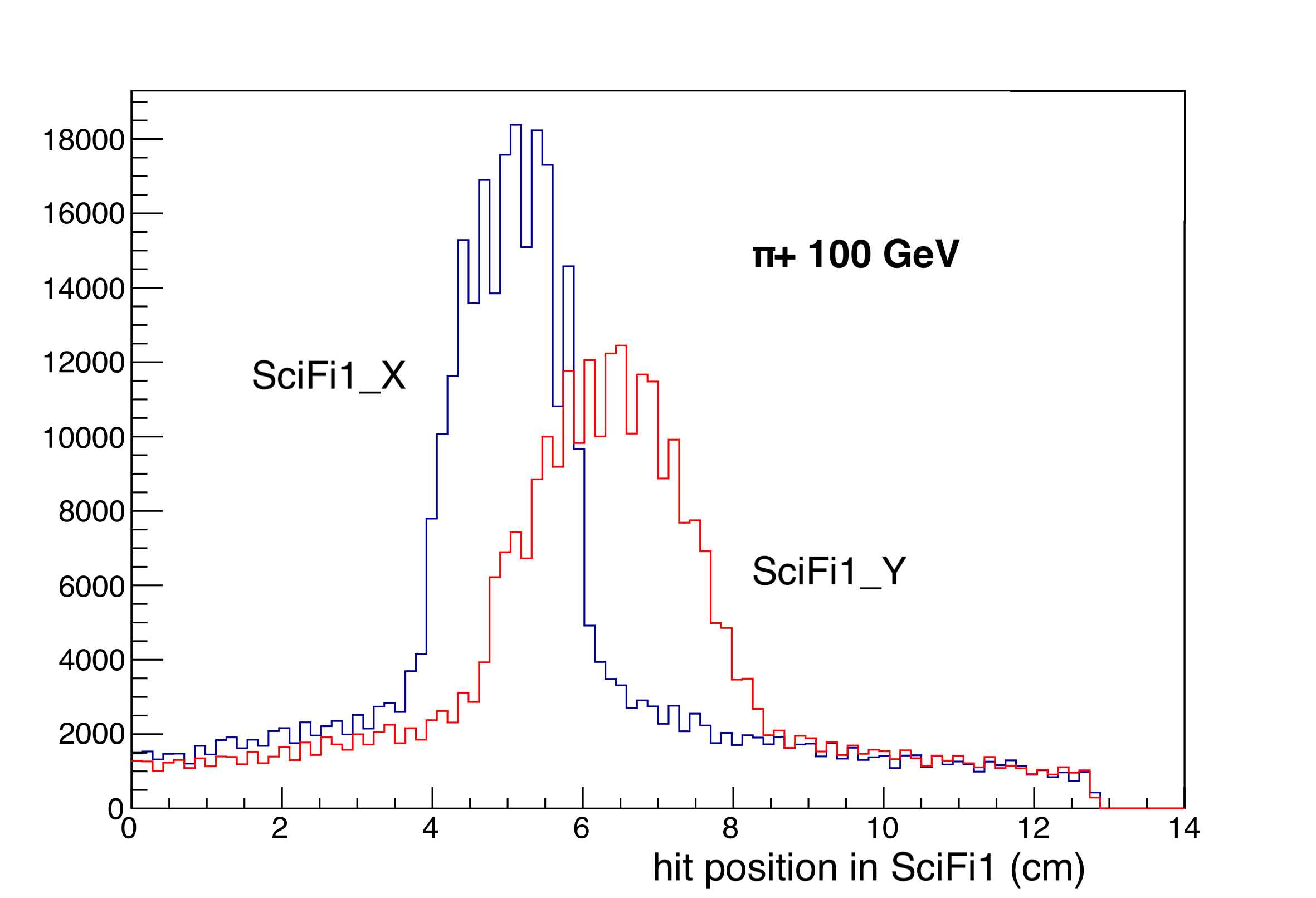}
\includegraphics[width=0.65\textwidth]{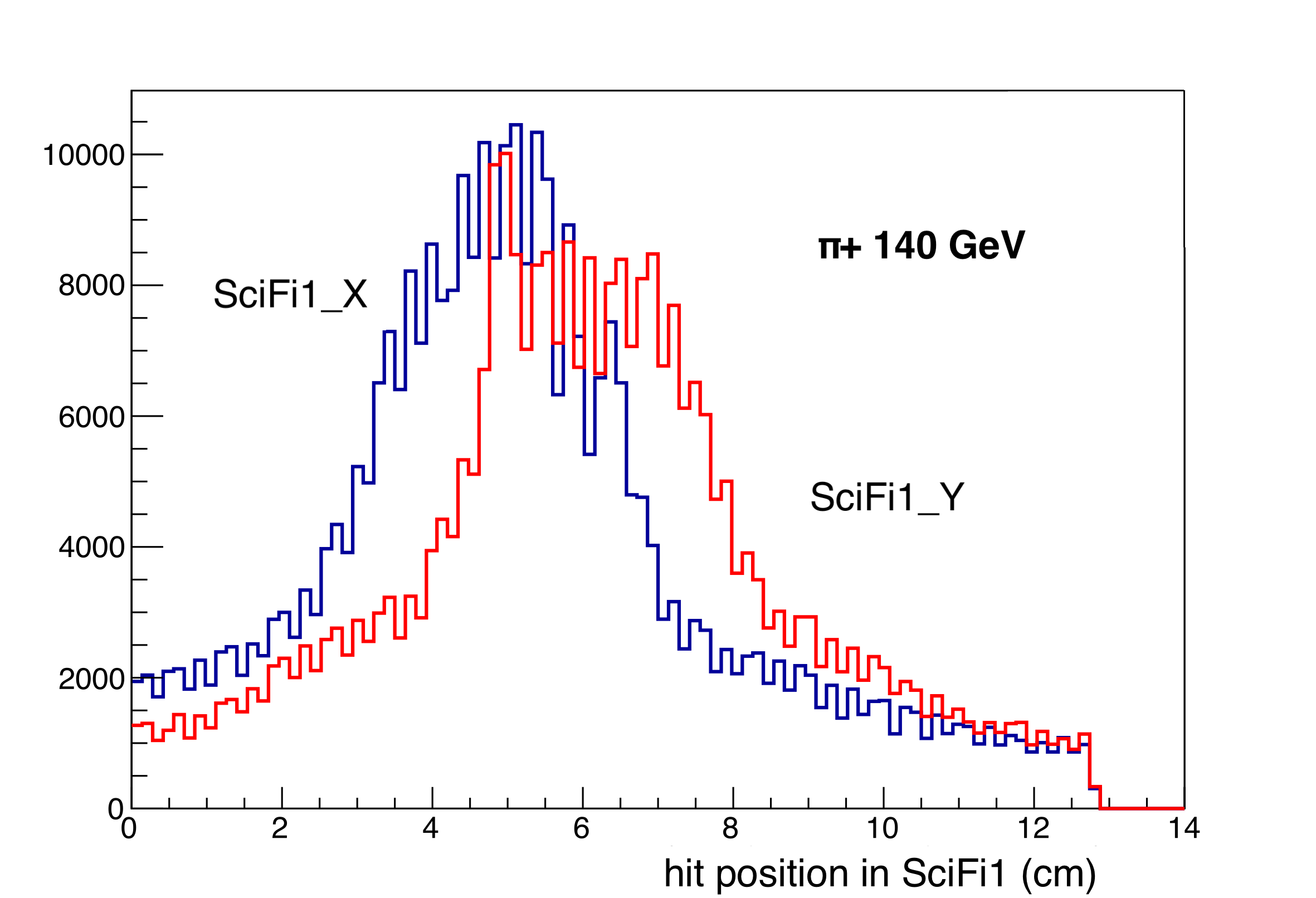}
\includegraphics[width=0.65\textwidth]{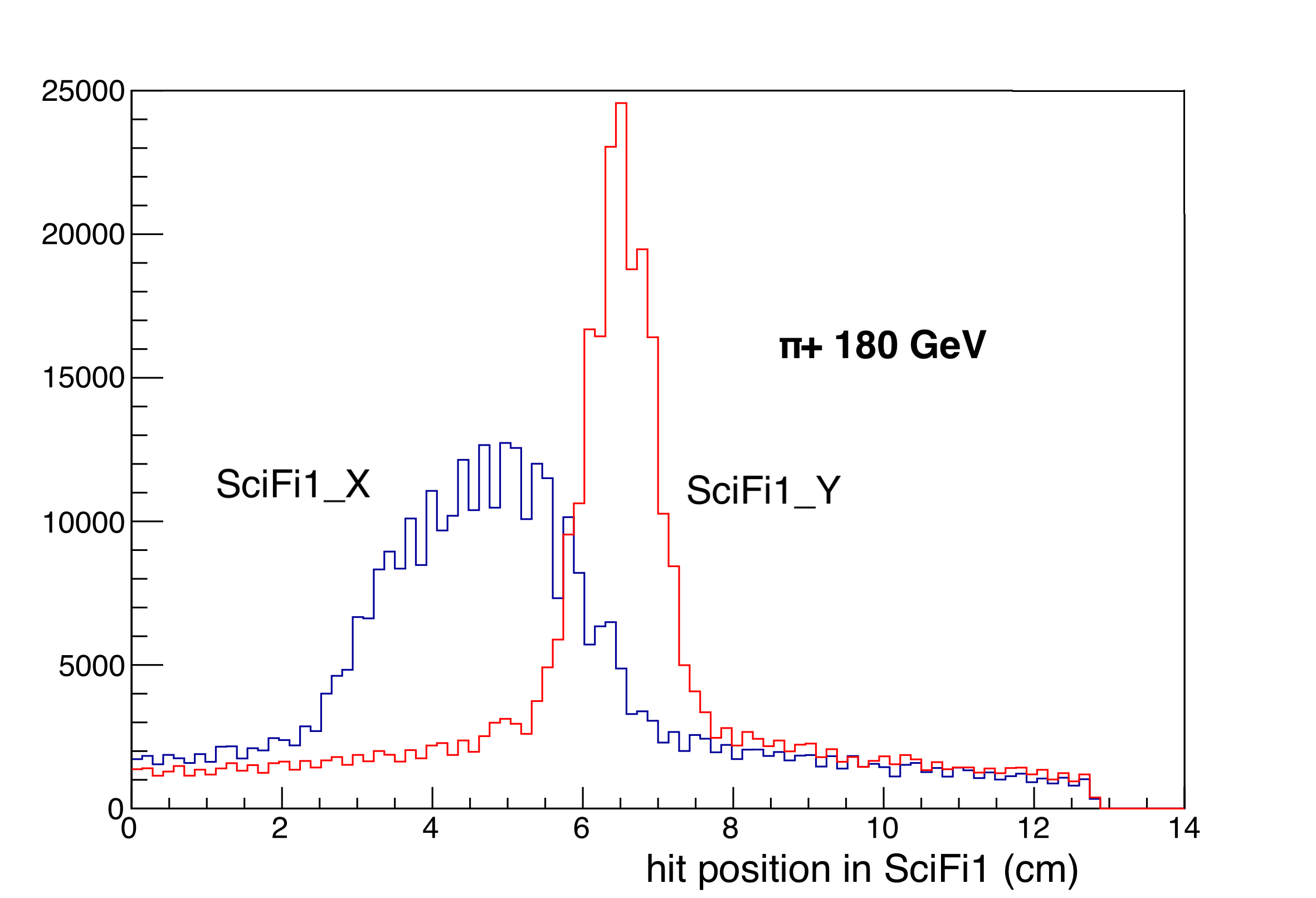}
\caption{ Beam profiles recorded in the most upstream SciFi 1 X and Y planes for positive hadron beams.}
\label{fig:BeamProfiles_positive}
\end{figure}

\begin{figure}[htbp]
\centering
\includegraphics[width=0.65\textwidth]{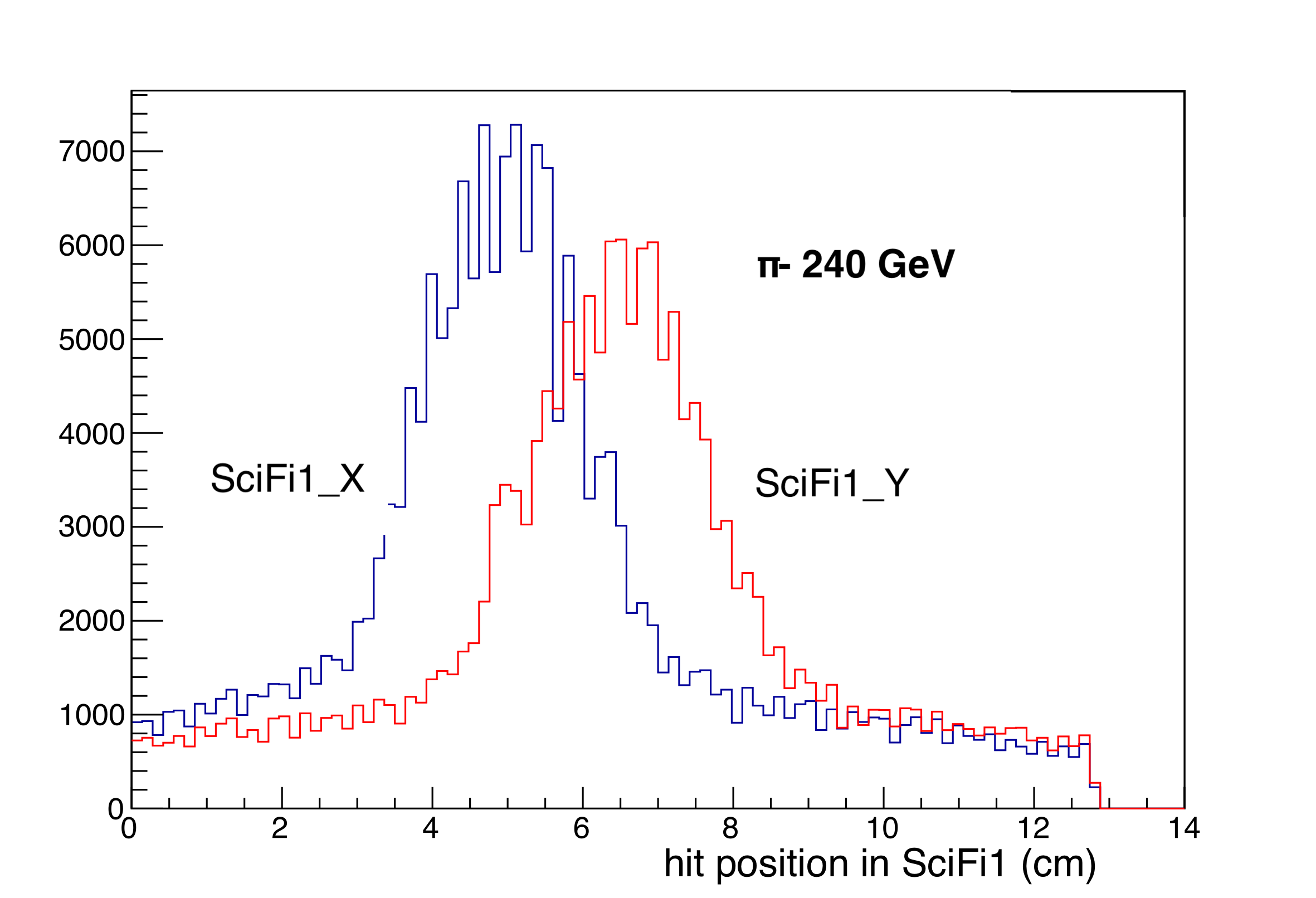}
\includegraphics[width=0.65\textwidth]{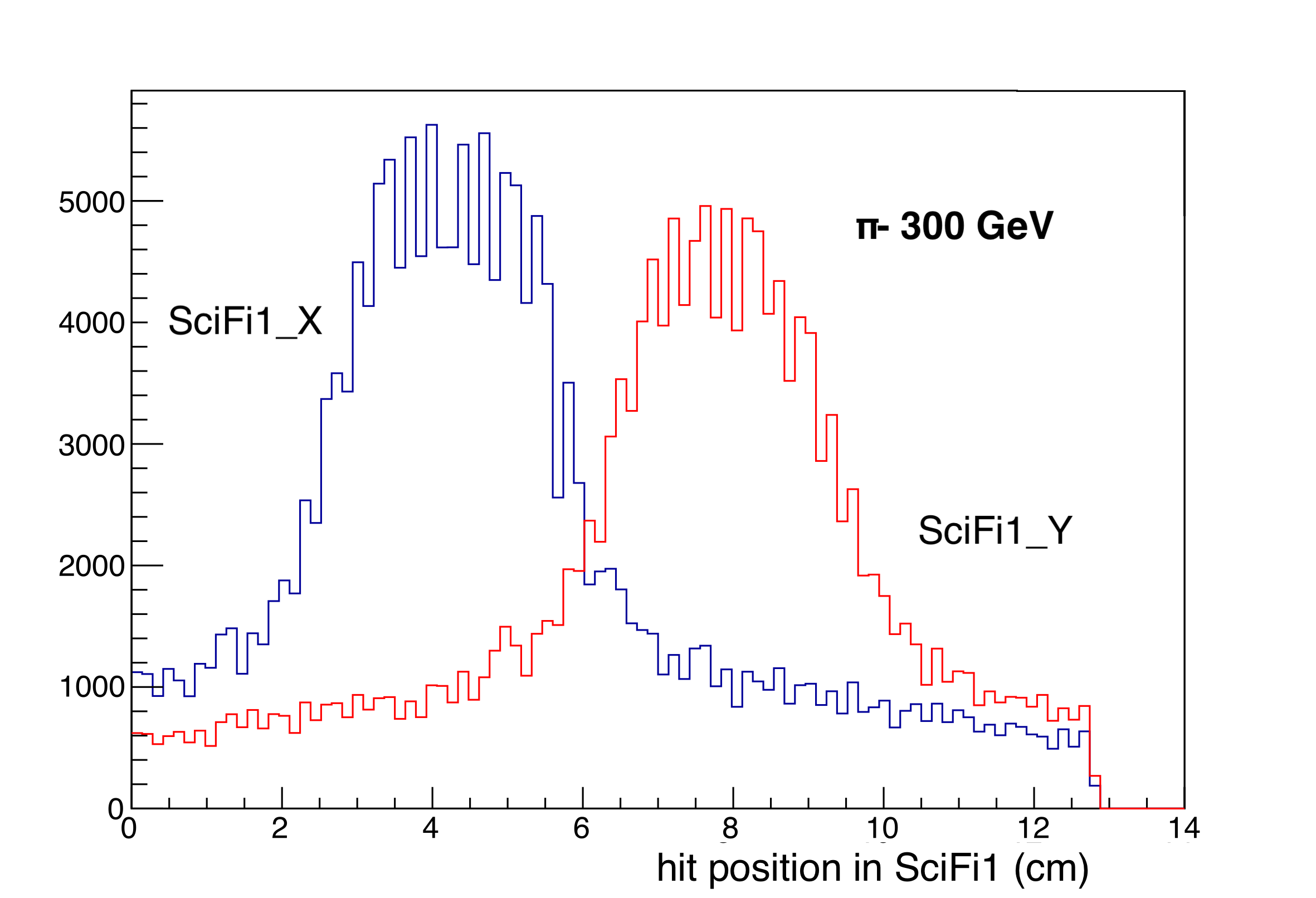}
\caption{ Beam profiles recorded in the most upstream SciFi 1 X and Y planes for negative pion beams. }
\label{fig:BeamProfiles_negative}
\end{figure}

The beam was aligned with the center of the detector. Figure~\ref{fig:BeamProfiles_positive} 
and 
~\ref{fig:BeamProfiles_negative}
show the beam profiles measured in the most upstream SciFi 1 planes X and Y for positive and negative hadron beams at the five energies. 

Data were stored for analysis with the following conditions:
\begin{itemize}
    \item since the task of the SciFi system was to tell precisely in which target wall a shower originated and there were four pairs of X,Y planes interleaving three target walls, the online filter asked that five out of eight planes had hits (example : 1x~1y~2y~3y~4y).
    \item since US noise was very low  (four SiPMs masked out of 800), all signals in HCAL were accepted, to guarantee no bias on shower shape analysis. 
\end{itemize}
The DAQ system, clone of the SND@LHC one, was able to collect up to $1.5~\times~10^7 $ events/hour.

Data were collected with three target configurations: a basic configuration with all three target walls, and configurations with one and two target walls, in order to be able to cross check the HCAL response dependence on the nuclear interaction depth in the target. 
Since the pion interaction length in iron is $20.4\ \rm cm$ and each target wall is $10\ \rm cm$ of Fe, in the 
three-walls configuration  sizable amounts of interactions occur also in the second and third wall (see table~\ref{tab:PionInteractionProb}), however the comparison with one- and two-walls configurations provides additional information on potential systematic errors in determining the hadron shower origin. In the one-wall configuration, the wall was between the third and the fourth Scifi stations; in the two-walls configuration, there was no wall in between the first and the second SciFi stations.
Note that over $20\%$ of pions 
cross all target walls without a hadronic interaction, even in the three-walls configuration, and end up interacting directly in the HCAL $20\ \rm cm$ thick iron walls.
\begin{table} [t]
\centering
\caption{ Fractions of pion interactions in consecutive $10\ \rm cm$ thick iron walls.}
\smallskip
\begin{tabular} { l  c c } 
\hline
 & interaction probability  & fraction of interactions   \\
\hline
first wall & $38.8\ \%$ & $50.4\ \%$   \\
second wall & $23.7\ \%$ & $30.8\ \%$  \\
third wall & $14.5\ \%$ & $18.8\ \%$   \\
\hline
total & $77\ \%$ & $100\ \%$   \\
\hline
\end{tabular}
\label{tab:PionInteractionProb}
\end{table}

The minimal sample of events to be stored for each configuration was set with a view to forthcoming analysis requirements.
The aim was that the HCAL energy calibration would contribute a negligible systematic error in measurements with the statistical uncertainty of the expected full sample  of neutrino interactions collected by SND@LHC in LHC Run3 (a few thousands, i.e. about $2\ \%$ uncertainty). A statistical precision of $0.1\ \%$ is necessary for uncovering and studying systematic uncertainties at the $1\ \%$ level.
In the configuration with three target walls, out of the events passing the filter conditions, the fraction featuring a hit coincidence between the X and Y planes of the most upstream station SciFi~1 was close to $10\ \%$ at 100~GeV and decreased going up in energy down to below $5\ \%$ at 300~GeV. 
Table~\ref{tab:TBData} shows the amount of data acquired in each of the fifteen configurations. 

In between changes of configuration, data collected with a broad muon beam ($3\times10^5$ muons/run), generated by closing the beam stopper upstream of the tested detector, were used to monitor the status of the SiPMs, and the stability of the peak position of the MIP signal checked.
A very high statistics run with a 160 GeV muon beam was also performed.


\section{Analysis}
\label{sec:analysis}

\subsection{Shower Tagging}
\label{subsec:ShowerTagging}

Since in the SND@LHC HCAL hadronic showers are observed downstream of an interaction target with about $2.5~\lambda_\text{INT}$  ($1.5~\lambda_\text{INT}$ in the test beam), the shower development measured in the US planes is highly dependent on the interaction origin in the target.
Therefore the first step in the analysis was to develop a shower tagging algorithm that, besides detecting the presence of a shower and providing a good separation between showering hadrons and muon background, estimated the shower origin point.

\begin{figure}[tb]
\centering
\includegraphics[width=0.45\textwidth]{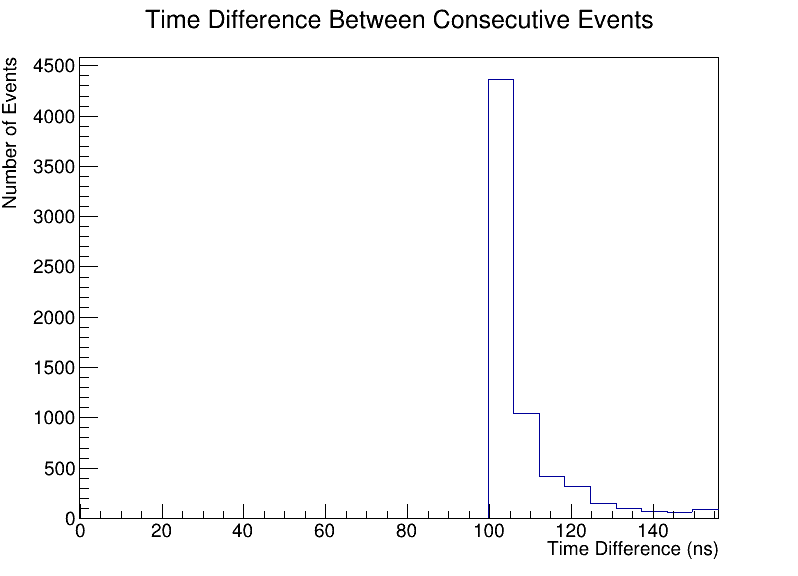}
\includegraphics[width=0.45\textwidth]{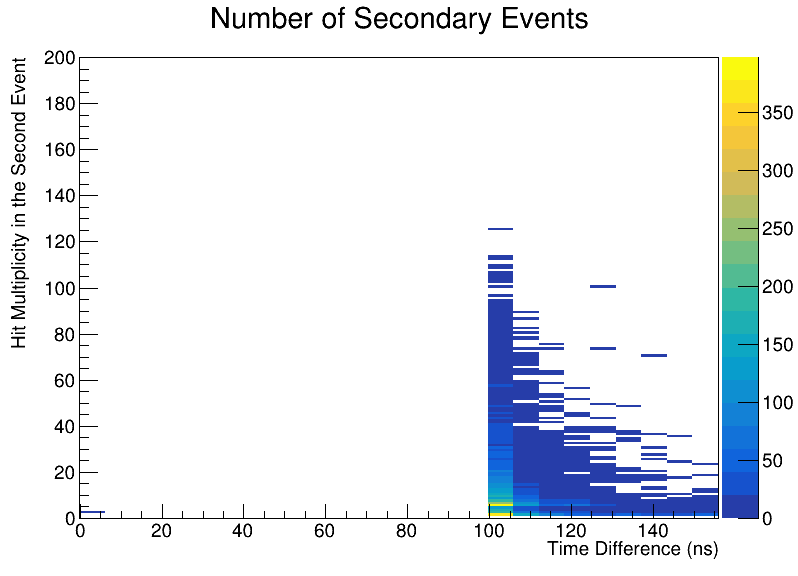}
\caption{Left: time difference between two consecutive events. Right: multiplicity of hits, both SciFi and US, in consecutive events as a function of the time difference between two consecutive events. 
}
\label{fig:consecutiveEvents}
\end{figure}

Preparatory to shower tagging was the selection of "in time" hits. In the LHC data throughout 2022 and 2023~\cite{Ringing1} a "ringing effect" was seen  where delayed hits and even new events
were created shortly after good events.
These events had the particular feature of having patterns of firing SiPMs correlated to the preceding good events.
Similarly in the test beam, in which, as mentioned in section~\ref{sec:detector}, events had a duration of $\approx 100$~ns, the data showed an increased amount of consecutive events that were only separated by approximately $100$~ns (figure~\ref{fig:consecutiveEvents}). This feature was investigated in more detail.

\begin{figure}[tb]
\centering
\includegraphics[width=0.6\textwidth]{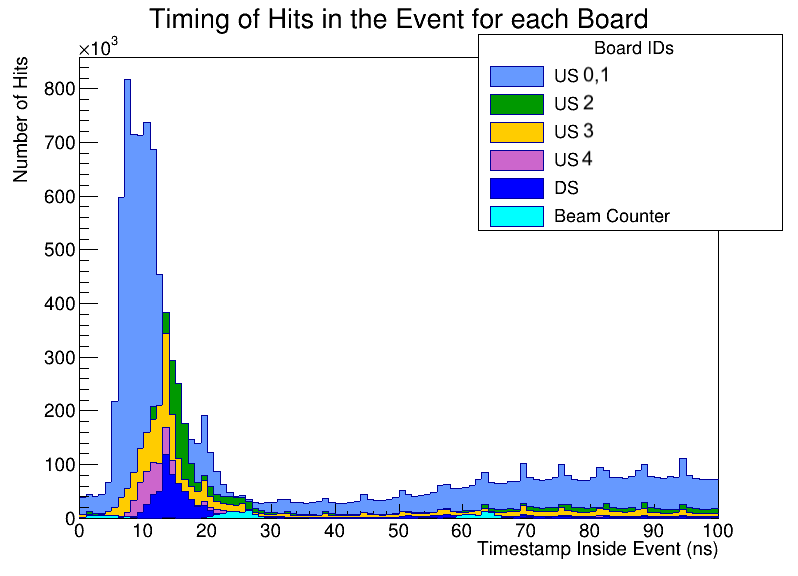}
\caption{Timestamp distribution of hits in 
the DAQ boards collecting HCAL and Beam Counter scintillators data. }
\label{fig:USHitsTimestamp}
\end{figure}

Figure~\ref{fig:USHitsTimestamp} shows the distribution of hit times within the $100$~ns event window 
in the HCAL DAQ boards.
Most hits tend to appear around coherent time-stamps for each respective DAQ board, forming clear peaks from $\approx 10$~ns to $\approx 20$~ns.
Since the DAQ board  collecting data from US0 is the same as for US1, its hit distribution comprises two overlapping peaks, slightly separated in time.
The distributions show an unexpected broad shoulder for times later than $30$~ns. This contribution continues even after the $\approx 100$~ns event time window closes, and as such can create new HCAL events immediately after the first one ends, leading to the excess seen in figure~\ref{fig:consecutiveEvents}. 

Noise events are mostly rejected by a filter condition applied to the SciFi data requiring that more than ten hits are detected within the whole SciFi system. Uncorrelated noise hits appear sporadically and sparsely, unlike hits from actual particles interacting in the detector. In most cases, the HCAL "ringing" events should not be capable of passing the SciFi filter. However, the hit time distributions show recurrent peaks that approximately match the $\approx 6.25$~ns clock cycles of the DAQ system, allowing a fraction of the "ringing" events to pass the selection.
In order to mitigate the effect, only events that do not have a preceding event within $150$~ns are considered. 

\begin{figure}[tb]
\centering
\includegraphics[width=0.6\textwidth]{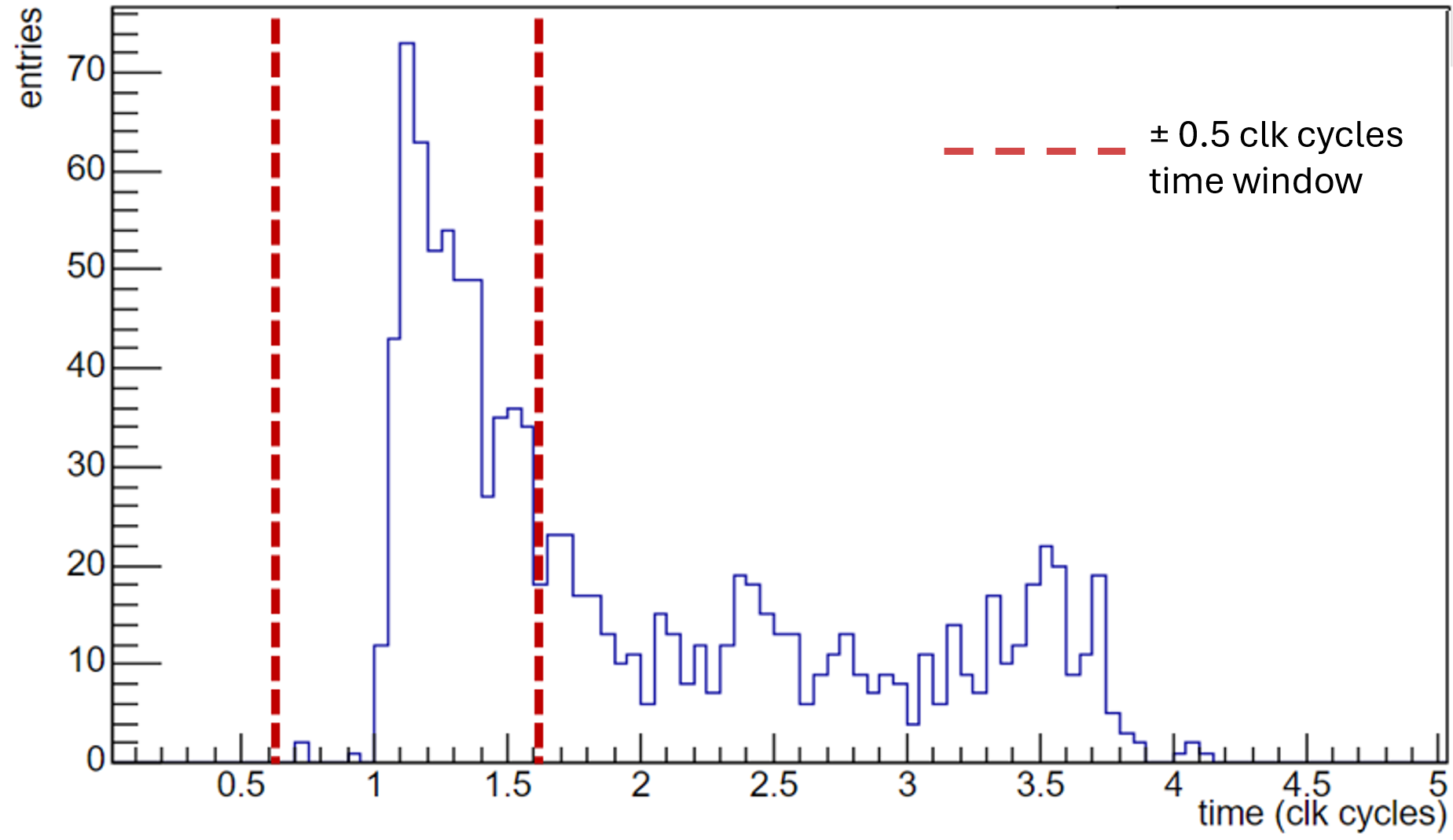}
\caption{Typical time distribution of SciFi hits in one event. SciFi hits are considered “in time" 
if they are within a $\pm0.5 $ clock cycle window around the most probable value of the SciFi hit time distribution (in this plot the range is bounded by the dashed lines). One clock cycle equals $\approx6.25$ ns.}
\label{fig:SciFi_tRef}
\end{figure}

\begin{figure}[hb]
\centering
\includegraphics[width=0.8\textwidth]{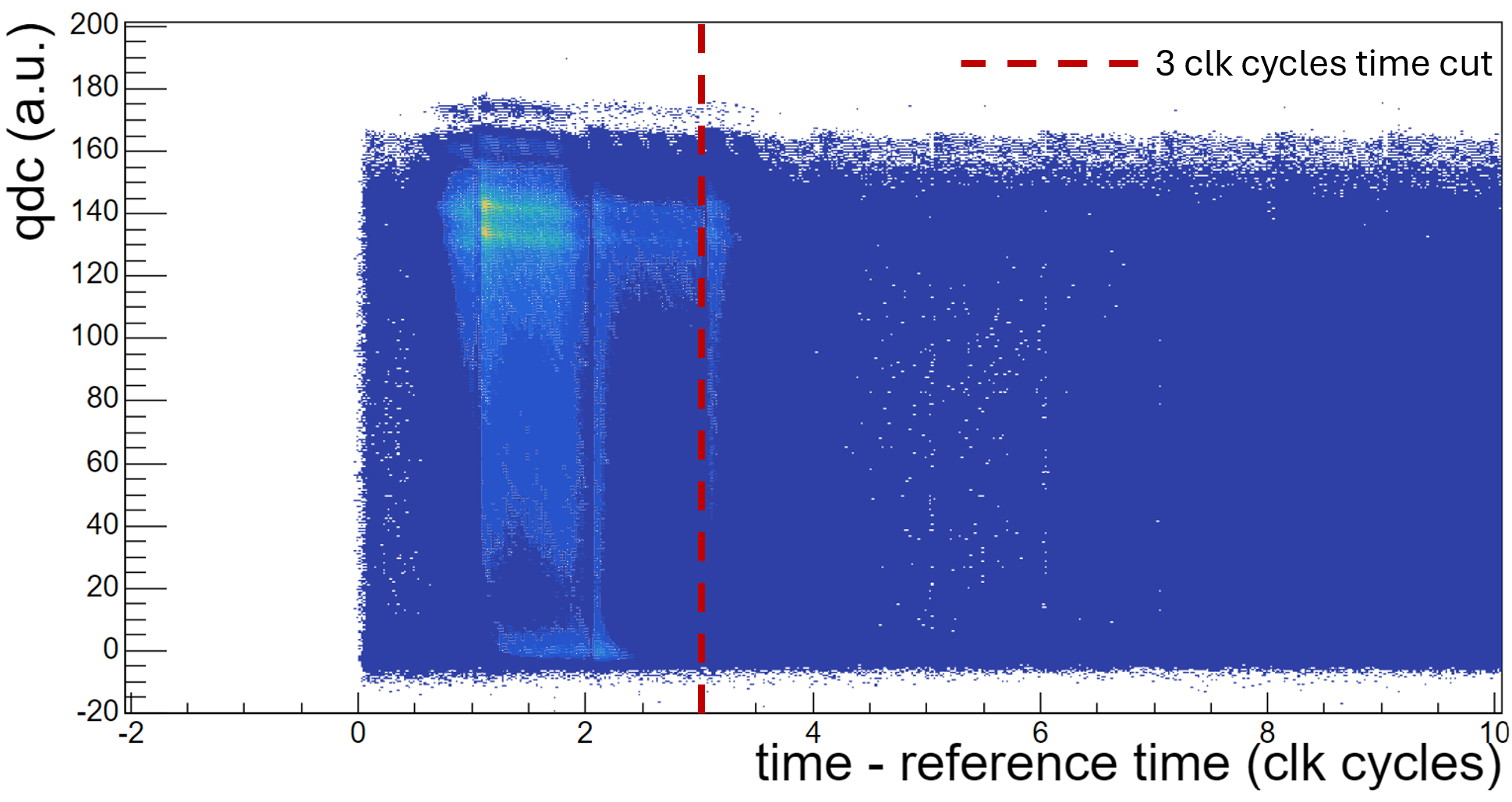}
\caption{Scatter plot of the digitized charge (QDC) versus time for all SiPM signals of US stations; the time is considered with respect to the SciFi reference time $t_{ref}$. US signals are considered “in time" if earlier than 3 clock cycles (in this plot the range is bounded on the right by the dashed line). One clock cycle equals $\approx6.25$ ns.}
\label{fig:US_tRef}
\end{figure}

Within an event, SciFi hits are considered “in time" if lying within $0.5$ clock cycles ($\simeq3$ ns) from the most probable value $t_{ref}$  of the SciFi hit distribution in time (figure~\ref{fig:SciFi_tRef}). 
US hits are considered “in time" if recorded within $+3$ clock cycles ($\simeq19$ ns) from $t_{ref}$. (figure~\ref{fig:US_tRef}). 
In the US stations, 
only hits registered by large SiPMs were considered, since signals from small SiPMs were found to be dominated by cross talk from the large SiPMs.

\begin{figure}[tb]
\centering
\includegraphics[width=1.0\textwidth]{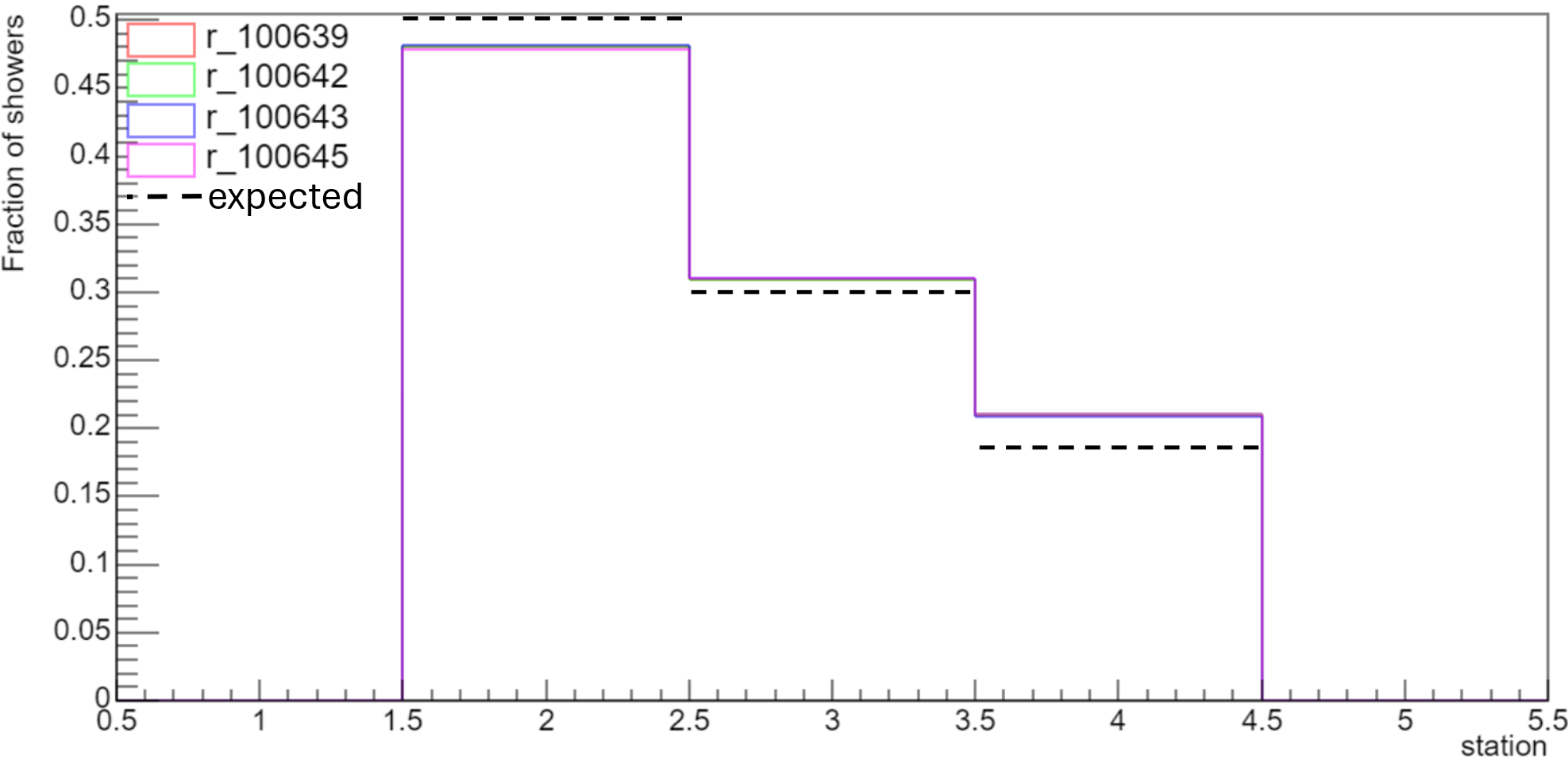}
\caption{Fraction of interactions in each SciFi station. Clearly, the first station does not tag any shower since it is placed upstream of the first iron wall. Four runs with the same nominal pion energy are shown; results are consistent with expectations in table~\ref{tab:PionInteractionProb}.}
\label{fig:showertag}
\end{figure}

The shower tagging algorithm only uses "in time" hits. Several algorithms were studied. The Shower Tagging via the Hit Density algorithm 
(section~\ref{subsubsec:HitDensityTagging}
was found the most effective. 
In this use, it is configured to search for "in time" hits in at least 35 consecutive channels (corresponding to about $0.9$~cm)  within a sliding window of 128 channels, along both the X and Y planes of each SciFi station in the target. The most upstream SciFi station satisfying the shower tagging requirement marks the iron wall where the shower originated. This tagging algorithm gave consistent results across different runs, energies, and number of iron walls installed in the target (figure~\ref{fig:showertag}).

\subsubsection{Shower Tagging through Hit Density}
\label{subsubsec:HitDensityTagging}

The hit density tagging based algorithm was developed by exploiting the test beam setup design, where the target walls between SciFi stations could be removed.
When no iron walls are present in the target, the SciFi hit distribution reproduces the beam particle density profile in all four SciFi stations. As iron walls are added, the beam shape becomes less distinct on the detectors downstream of the walls due to the showers produced by the interacting particles.

Since only a fraction of particles from the incident beam is going to interact in each target wall, the beam shape seen when no iron walls are in the target upstream should be recoverable by properly selecting the events where a shower has not yet started.

Thus, if a criterion such as the density of hits within a specific window of channels is defined, an optimal threshold can be derived by comparing the similarity between the beam profile without an iron target upstream of the SciFi plane in question, and the beam profile when selecting events that have not produced a shower upstream of the selected plane, with a fully instrumented target.

This analysis was done using the default timing window for the SciFi hits of [-0.5;~+0.5] clock cycles from $t_{ref}$, 
i.e. from the most probable value of the SciFi hit distribution in time, as well as an alternative larger acceptance with a window for "in time" events of [-0.5;~+2.0] clock cycles,
to mirror the constraints to be used in upcoming muon neutrino analyses on behalf of the SND@LHC collaboration. 
Unless stated otherwise, all the plots shown will be done using the default timing window.

The beam profiles were obtained from the fired SiPM channels in the second SciFi station. The analysis  compared data from runs with three target walls, with runs using only two target walls, as the latter provide beam profiles without shower for the second SciFi station.

\begin{figure}[tb]
\centering
\includegraphics[width=0.45\textwidth]{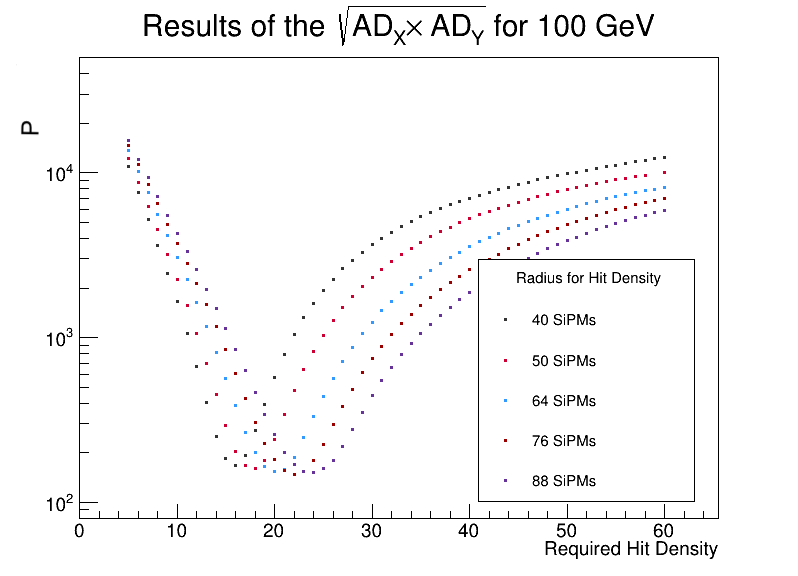}
\includegraphics[width=0.45\textwidth]{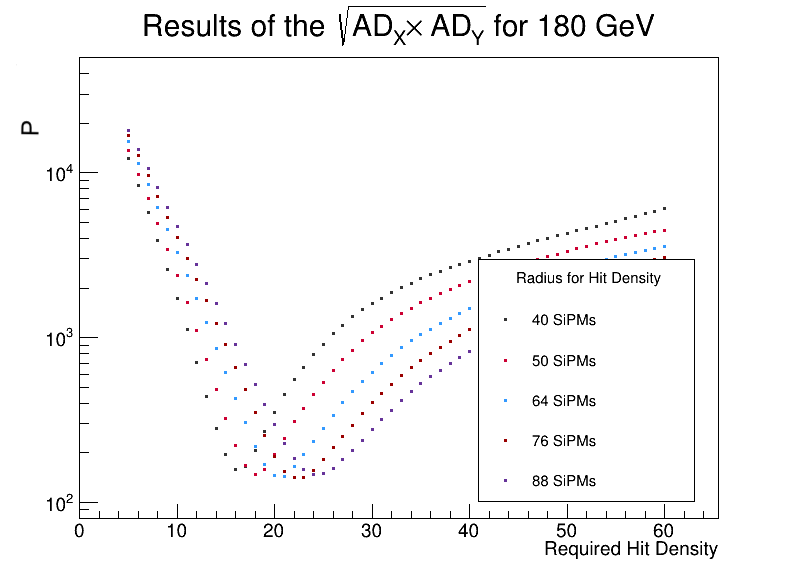}
\caption{Result of the performance metric P of the Hit Density tagging algorithm as a function of the hit density and of the sliding window Radius, for 100~GeV (left) and 180~GeV (right) hadron beams.}
\label{fig:TestResults_100677}
\end{figure}

In order to evaluate the similarity between profiles, both the $\chi^2$ and Anderson-Darling (AD)~\cite{AndersonDarling}
methods were used. The AD method was chosen as the baseline method for comparison. Comparisons were made between the X profiles and the Y profiles, and a performance metric that combined both profiles was chosen as:
\begin{equation}
\label{eq:AD2}
\begin{aligned}
\text{P} &= \sqrt{\text{AD}_X \times \text{AD}_Y}
\\
\end{aligned}
\end{equation} 
where P is the performance metric we are trying to minimize and AD$_{X,Y}$ are the results for the AD test in the respective X,~Y profiles. A similar measurement was done for the $\chi^2$ method.
The hit density, defined as the number of "in-time" fired channels within a set range of channels, required to tag a shower was varied from 5 to 60 hits within sliding windows that varied from 10 channels to 176 channels, i.e. radii (R) between 5 and 88 channels.
The performance  P as a function of the hit density and of the radius of the sliding window can be seen in figure~\ref{fig:TestResults_100677} for 100 and 180~GeV data.
P is minimized in a broad range for hit densities  between 15 and 35 in sliding windows with Radii larger than 40. 
The broadness of the minimum is related to the variability of development among showers, since the shower origin can be anywhere along the depth of the wall preceding the second SciFi station.
To ensure that showers are enough developed when sampled by SciFi, we choose a Hit Density working point of 35 hits in a window of R=64 half width.

\begin{figure}[tb]
\centering
\includegraphics[width=0.5\textwidth]{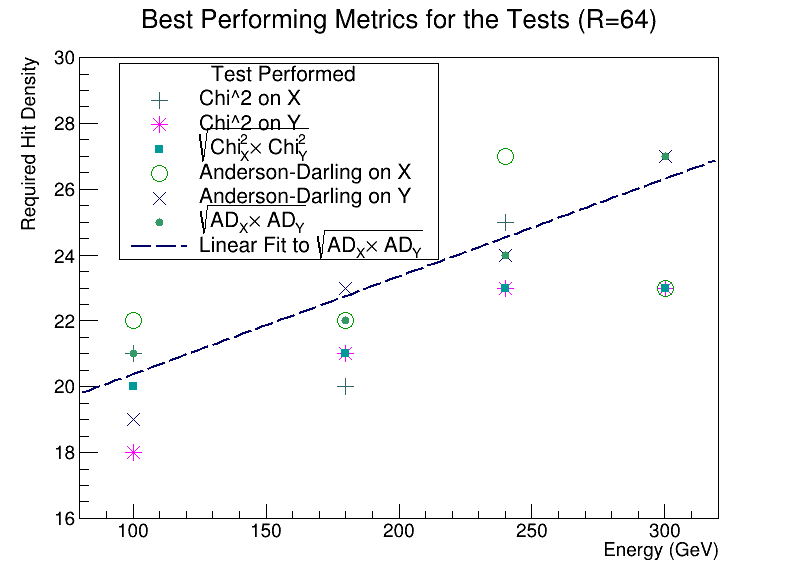}
\includegraphics[width=0.5\textwidth]{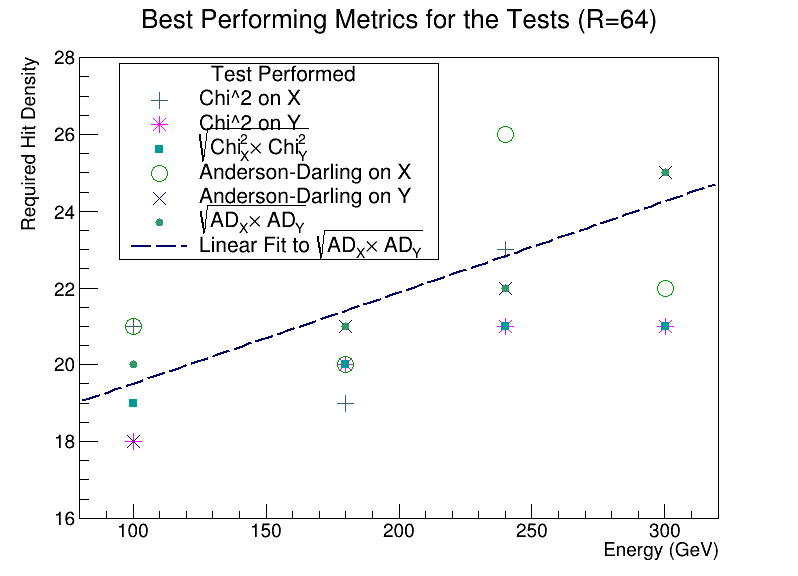}
\caption{Summary of hit densities that minimize
 the metric P as a function of the beam energy, for a sliding window with a radius R=64. 
Top figure: SciFi timing window of [-0.5; +0.5] clock cycles. Bottom figure: SciFi timing window of [-0.5; +2.0] clock cycles.}
\label{fig:BestTestResults_Density}
\end{figure}

Since the number of particles in the shower increases with the logarithm of the parent hadron energy,  we do expect a small bias of the algorithm with energy.
To quantify the size of this effect,
the method was applied to all available energies, as shown in figure~\ref{fig:BestTestResults_Density}. 
A very small energy dependence can be seen, 
for both the default SciFi timing window and the alternative one, however the effect is well within the broadness of the minimum of the performance metric P in
figure~\ref{fig:TestResults_100677}
and can be neglected.

\begin{figure}[tb]
\centering
\includegraphics[width=0.6\textwidth]{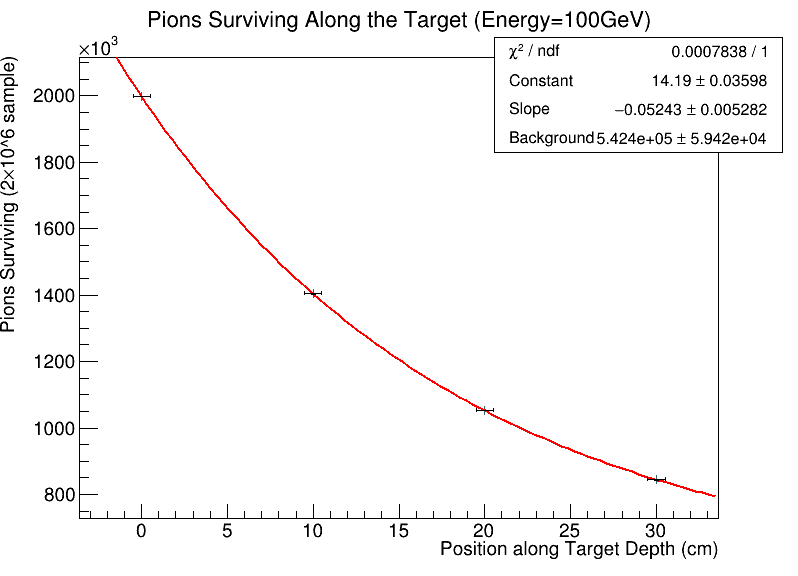}
\caption{Distribution of surviving pions along the target depth in a $2\times10^6$ sample at 100~GeV in the three-wall target configuration. The red line is the fit using equation~\eqref{eq:PionFit}; Slope stands for 
$-1/\lambda_\text{INT}$.}
\label{fig:SurvivingPions100GeV}
\end{figure}

Finally,
we checked that, using the shower tagging algorithm, the measured interaction length of pions in iron was coherent with the PDG value ($\lambda_\text{INT})=$20.42~cm~\cite{10.1093/ptep/ptaa104}). 
We counted how many pions were found to interact in each of the three target walls. The surviving pions ($SP$) along the target will be distributed as:
\begin{equation}
\label{eq:PionFit}
\begin{aligned}
SP &= \text{Background}+e^{\text{constant}-z/\lambda_\text{INT}}
\\
\end{aligned}
\end{equation}
where Background is the muon contamination in the pion beam. 
A summary of the estimated interaction lengths  can be seen in table~\ref{tab:InteractionLength}, and an example fit in figure~\ref{fig:SurvivingPions100GeV}. All the measurements
are consistent with the expected $\lambda_\text{INT}$ of pions in iron.
\begin{table}
\centering
\caption{ Pion interaction length in iron  derived by tagging showers using the Hit Density method.
}
\smallskip
\begin{tabular}{ c |c c  }
    \hline
    Energy & \multicolumn{2}{ c }{estimated interaction length (cm)} \\ \cline{2-3}
    (GeV)& [-0.5;+0.5] (clock cycles) & [-0.5;+2.0] (clock cycles) \\ \hline 
    100 & $19.2\pm_{1.7}^{2.1}$ & $18.5\pm_{1.6}^{1.9}$ \\ 
    140 & $20.0\pm_{1.2}^{2.2}$ & $19.6\pm_{1.7}^{2.1}$\\ 
    180 & $18.2\pm_{1.5}^{1.8}$ & $17.9\pm_{1.5}^{1.7}$\\ 
    240 & $20.4\pm_{1.9}^{2.3}$ & $20.0\pm_{1.2}^{2.2}$\\ 
    300 & $20.0\pm_{1.2}^{2.2}$ & $19.6\pm_{1.7}^{2.1}$\\
    \hline
\end{tabular}
\label{tab:InteractionLength}
\end{table}

\subsection{Energy Calibration Method}
\label{subsec:HadronicEnergyMeasurement}

The data samples collected in the target configuration with three walls (section~\ref{subsec:data}) were split into “calibration" and “test" samples. For every event, the main quantities considered were: origin of shower as measured with the shower tagging algorithm, the digitized integral charge (QDC) of the SiPM signals summed over “in time" SciFi hits ($QDC_{SciFi}$) and QDC sum of “in time" US hits ($QDC_{US}$). 

\begin{figure}[tb]
\centering
\includegraphics[width=0.8\textwidth]{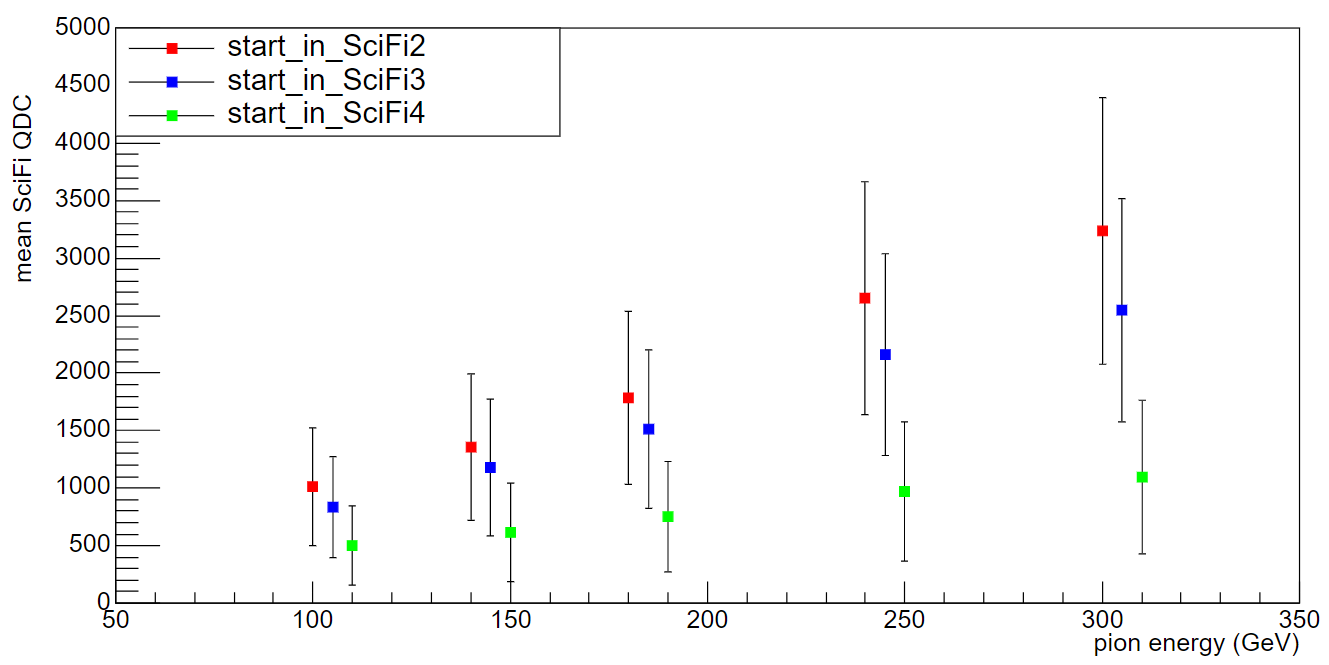}
\includegraphics[width=0.8\textwidth]{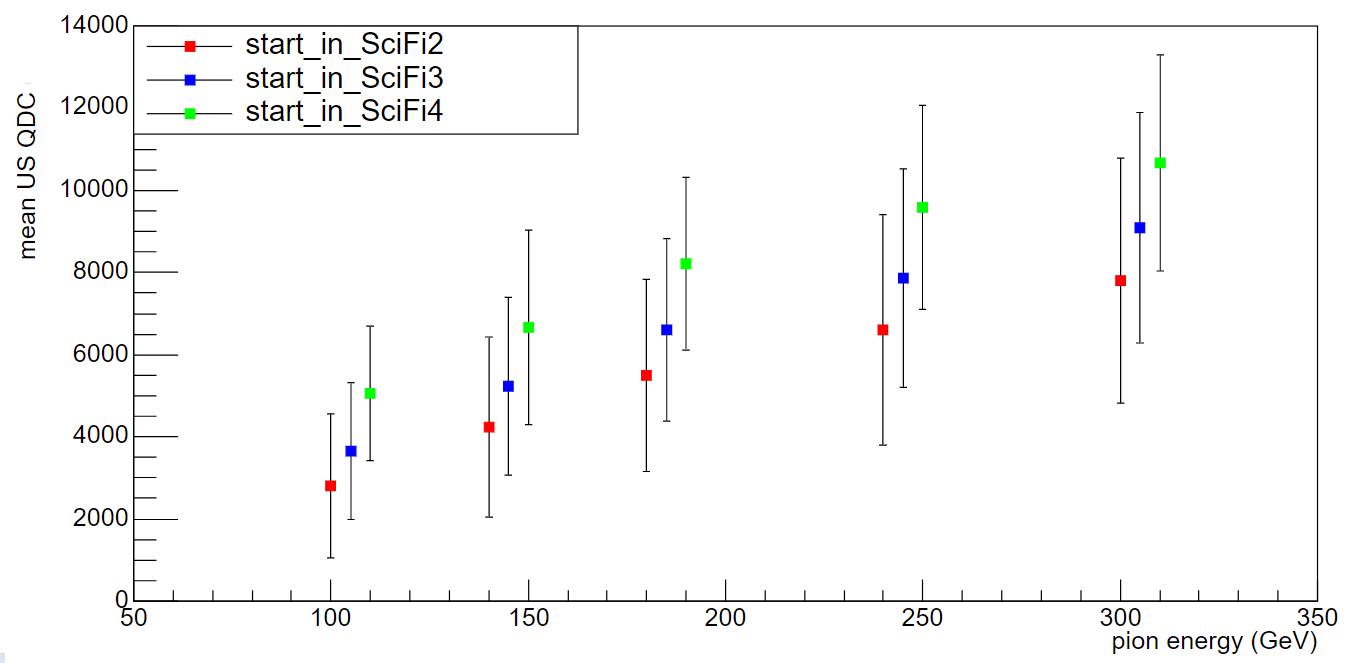}
\caption{ Top figure: mean $QDC_{SciFi}$ observed in events with different beam energies and start of shower. The error bars represent the standard deviations of the corresponding $QDC_{SciFi}$ distributions. As expected, the mean $QDC_{SciFi}$ increases with the beam energy but it decreases the more downstream the shower begins.
Bottom figure: mean $QDC_{US}$ observed in events with different beam energies and start of shower. The error bars represent the standard deviations of the corresponding $QDC_{US}$ distributions. As expected, the mean $QDC_{US}$ increases with the beam energy and also grows the more downstream the shower originates.}
\label{fig:scifi_vs_us_vs_energy}
\end{figure}

As shown in figure~\ref{fig:scifi_vs_us_vs_energy},
both the mean $QDC_{SciFi}$ and $QDC_{US}$ increase with beam energy. 
However, while the mean $QDC_{SciFi}$ decreases the more downstream the shower originated along the target, 
the mean $QDC_{US}$ increases the more downstream the shower started
within the target.
This behaviour is expected and can be modeled as:
\begin{equation}
\label{eq:Eline}
\begin{aligned}
Etot & = k \times QDC_{SciFi} + \alpha \times QDC_{US}
\\
\end{aligned}
\end{equation} 
where $Etot$ (GeV) is the hadron energy, $k$ (GeV/QDC) and $\alpha$ (GeV/QDC) two calibration constants to be determined.

\begin{figure}[tb]
\centering
\includegraphics[width=1.0\textwidth]{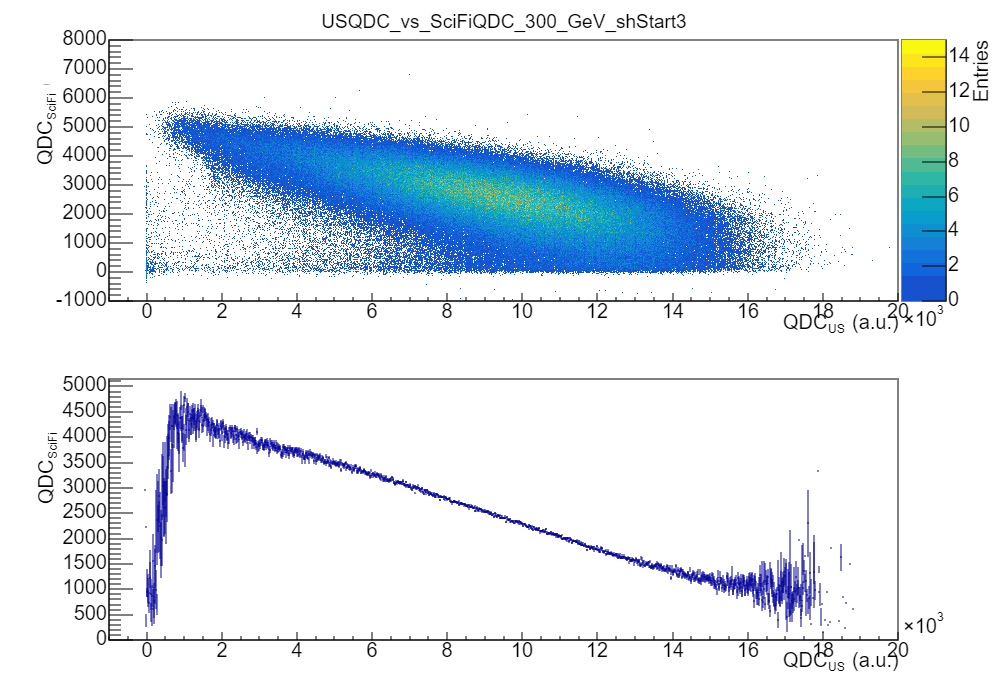}
\caption{Top: $QDC_{SciFi}$ vs $QDC_{US}$ scatter plot for 300 GeV beam energy and shower origins tagged in SciFi2.  Bottom: the corresponding profile along the X axis.
}
\label{fig:prof_sh3}
\end{figure}

The data are further organized into calibration sub-samples representing all 15 combinations of beam energy and iron wall where the shower initiated. In each sub-sample, the event distribution in the $QDC_{SciFi}$ and $QDC_{US}$ phase space is studied and a clear linear anti-correlation observed 
(figure~\ref{fig:prof_sh3}). 
Then, either a linear fit or a  Principal Component Analysis (PCA)~\cite{PCA}
is used to calculate $k$ and $\alpha$ of eq.\eqref{eq:Eline}.
The PCA is an iterative event-by-event procedure that finds the axis with respect to which the dispersion of events is minimized (the major axis in an elliptic phase space). 
The fit instead is performed globally on the scatter plots like in the upper plot in figure~\ref{fig:prof_sh3},
and similar for other energies and data samples.
The $QDC_{US}$ range for performing the fit is chosen corresponding to the linear range of the scatter plot profiles, as shown in the lower plot in figure~\ref{fig:prof_sh3}.
From the line describing the major axis of the phase space ellipse one extracts k and $\alpha$, while the spread of the distribution along the minor axis provides an estimate of the measurement resolution.

\begin{figure}[tb]
\centering
\includegraphics[width=1.00\textwidth]{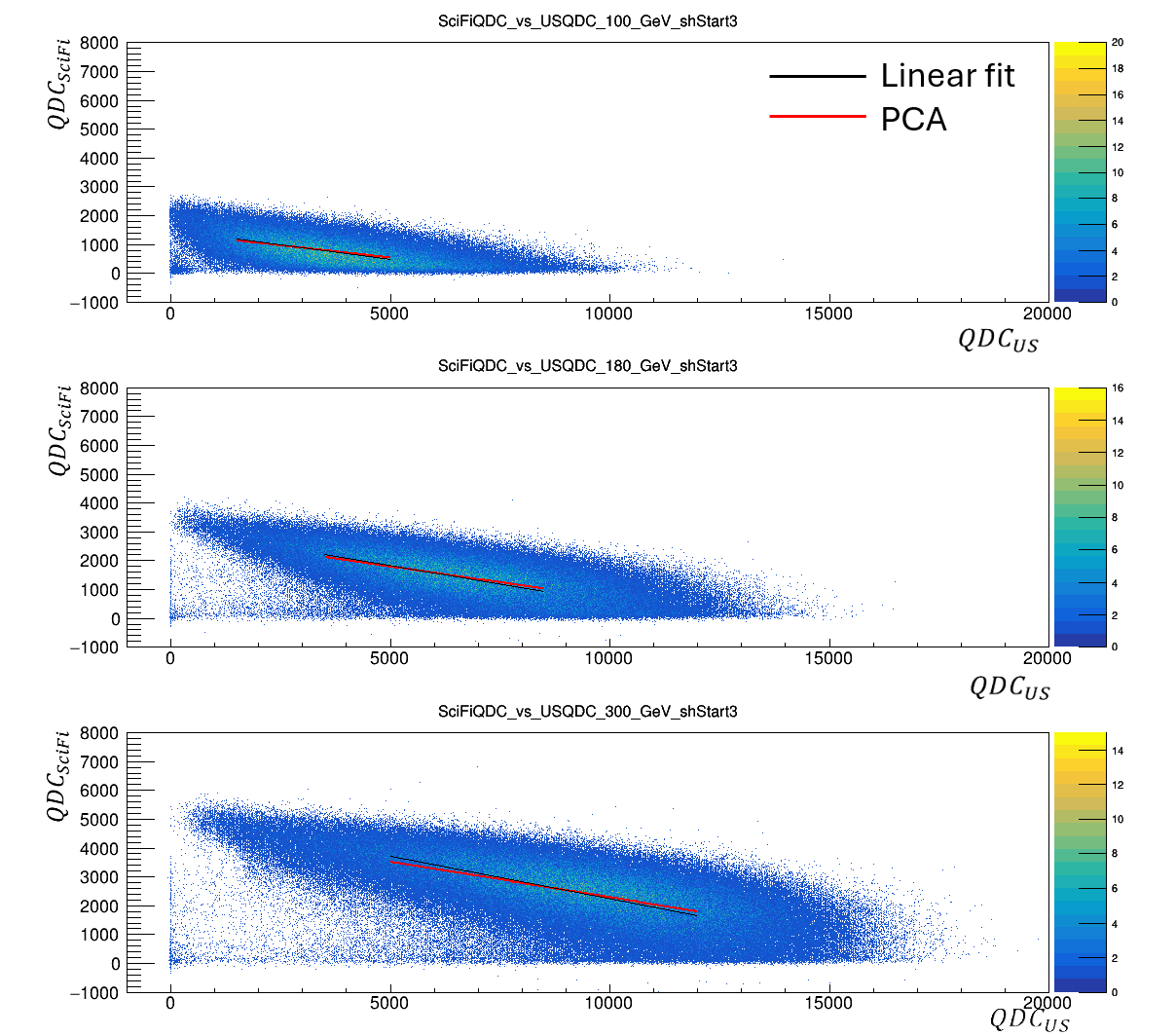}
\caption{Comparison between linear fit and PCA for finding the major axis of the 2D QDC correlation ellipses, used to evaluate the calibration constants $k$ and $\alpha$. The plots show data from 100, 180 and 300 GeV beam energy for showers originated in target wall 2 (tagged by SciFi3).}
\label{fig:PCA_shStart_3}
\end{figure}

Figure~\ref{fig:PCA_shStart_3} 
shows that 
the $QDC$ distributions exhibit a good correlation of SciFi and US when the shower starts in the first or the second iron wall ("ordinary" showers), and both the fit and PCA methods are able to find the major axes of the ellipses.
The correlation is much less pronounced when the shower begins "late" in the final section
of the target, the third iron wall in the test beam setup: in this configuration the shower is sampled by one SciFi station only (SciFi4) in the target, and the energy deposit is maximal in the first US layer, which can induce saturation of the SiPMs, as discussed later on. 
In the following, ordinary and late showers are discussed separately.

\subsection{Energy Measurement for  Ordinary Showers}
\begin{figure}[tb]
\centering
\includegraphics[width=0.65\textwidth]{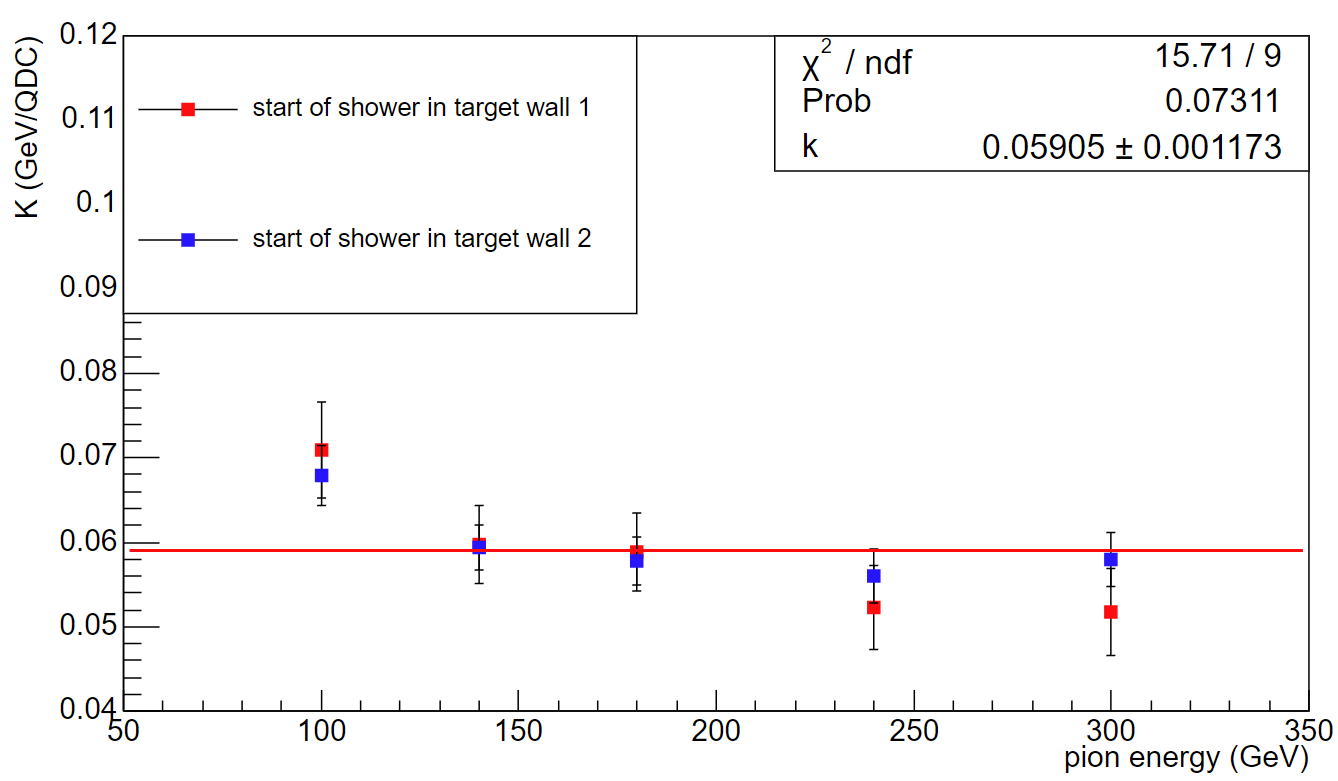}
\caption{$k$ values extrapolated from PCA, for each combination of beam energy and start of shower. 
}
\label{fig:k_vs_energy}
\end{figure}

\begin{figure}[tb]
\centering
\includegraphics[width=0.65\textwidth]{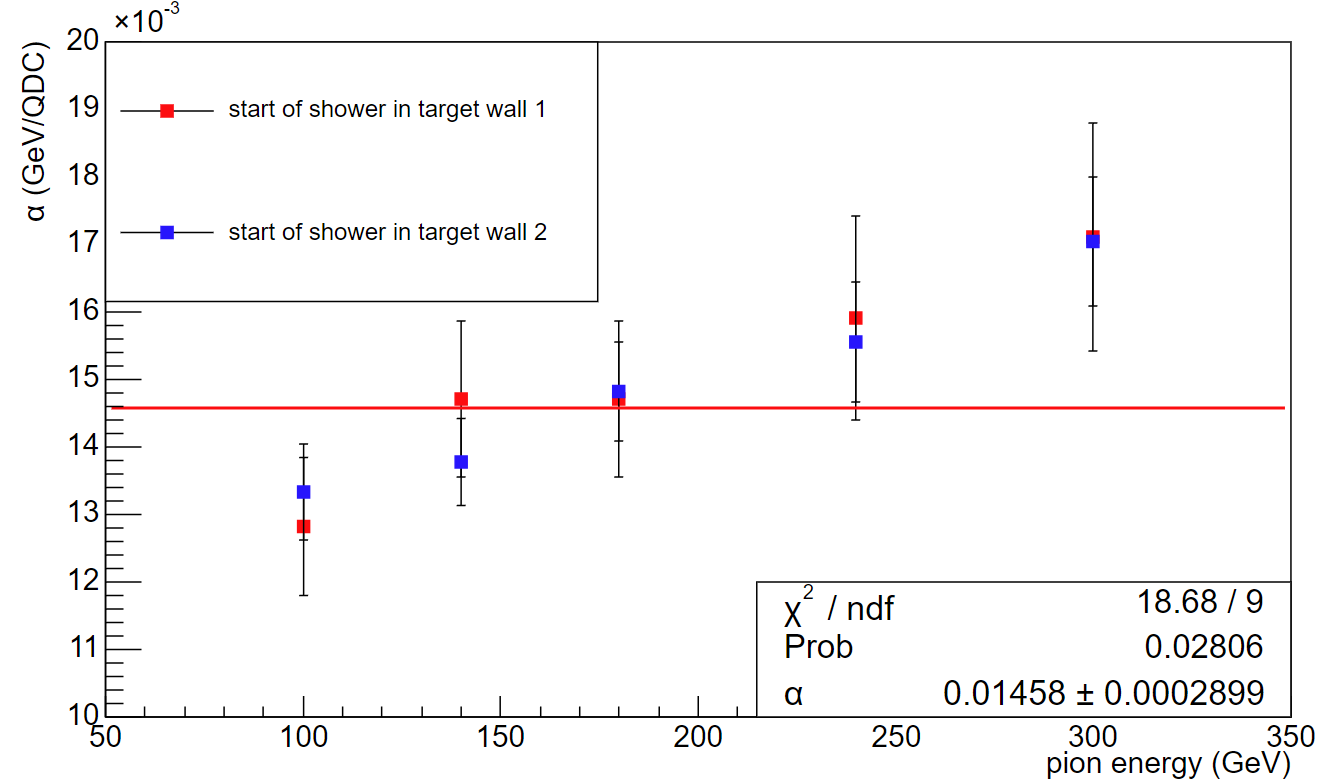}
\caption{$\alpha$ values extrapolated from PCA, for each combination of beam energy and start of shower. 
}
\label{fig:alpha_vs_energy}
\end{figure}

Ordinary showers are at a stage of development that allows for sampling in the target by at least two SciFi stations in both X and Y projections. 
They originate in walls 1 and 2 in the test beam setup.
The $k$ and $\alpha$ constants are calculated from the ellipse center and the axis direction of the $QDC_{SciFi}$ versus $QDC_{US}$ distributions, figure~\ref{fig:k_vs_energy} and figure~\ref{fig:alpha_vs_energy}.
The resulting average calibration constants are $k=(0.059 \pm 0.006)$ GeV/QDC and $\alpha=(0.0145 \pm 0.0010)$ GeV/QDC, 
where the errors include a $\pm5~\%$ systematic uncertainty for the slight variation with energy observed over the tested range.

\begin{figure}[tb]
\centering
\includegraphics[width=0.65\textwidth]{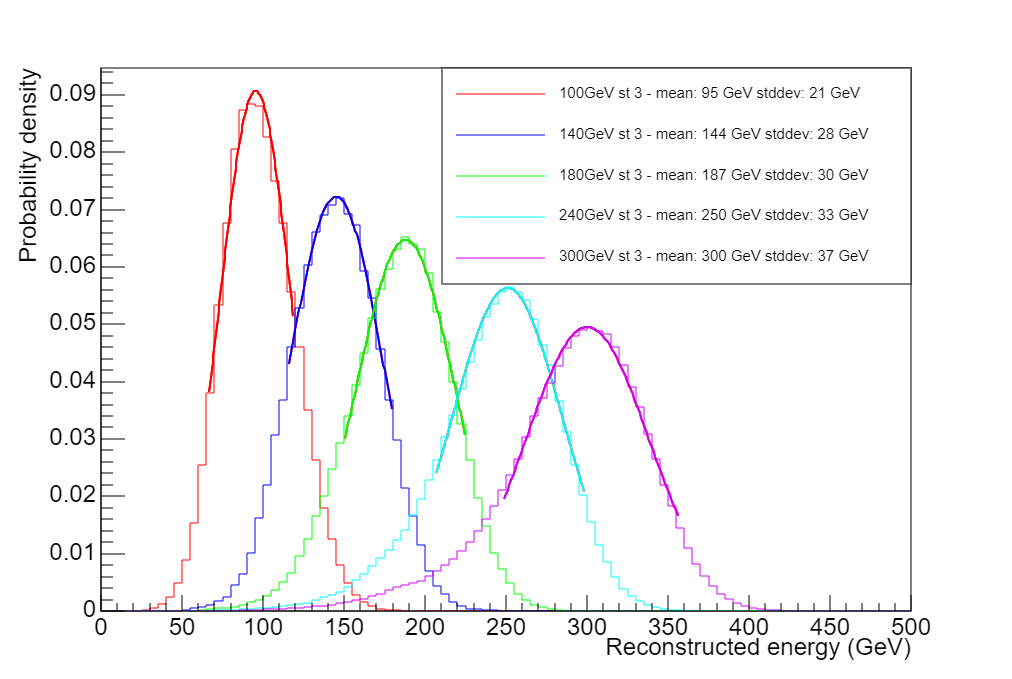}
\caption{Reconstructed energy $Etot$ 
for showers starting in SciFi3, with different beam energies. 
}
\label{fig:Ereco_st3}
\end{figure}

\begin{figure}[tb]
\centering
\includegraphics[width=0.55\textwidth]{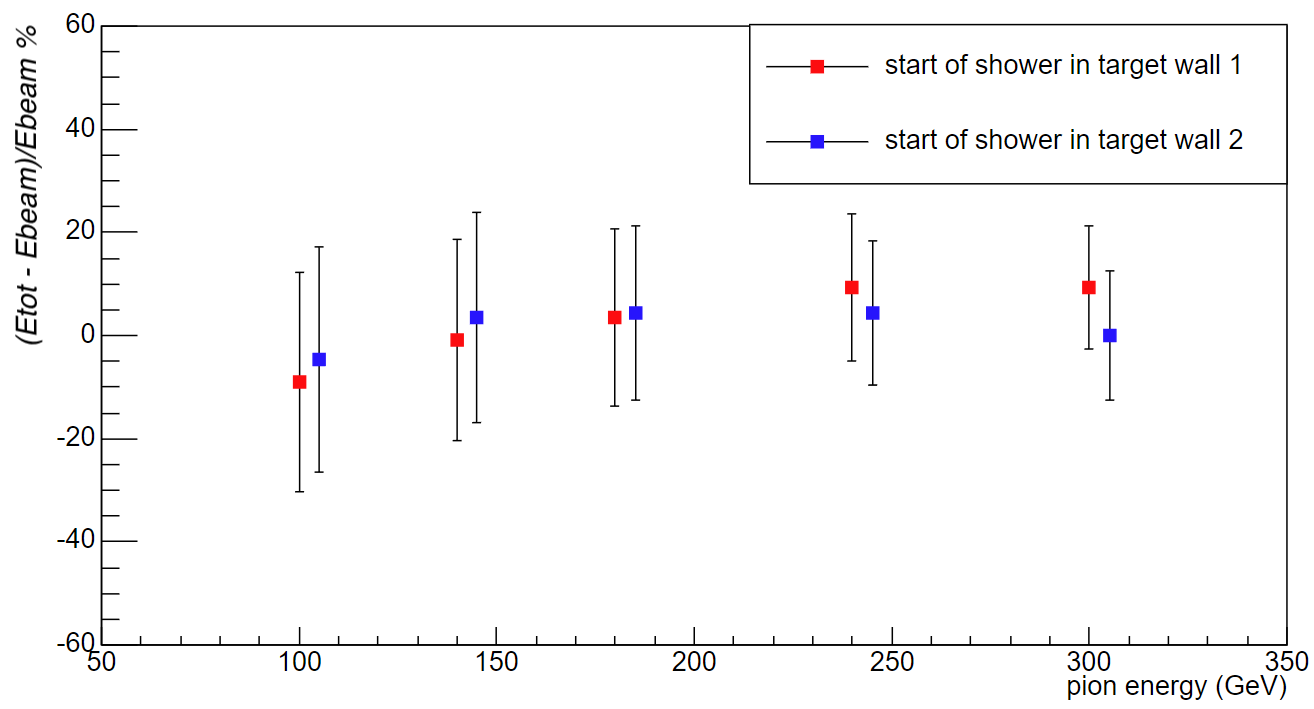}
\includegraphics[width=0.55\textwidth]{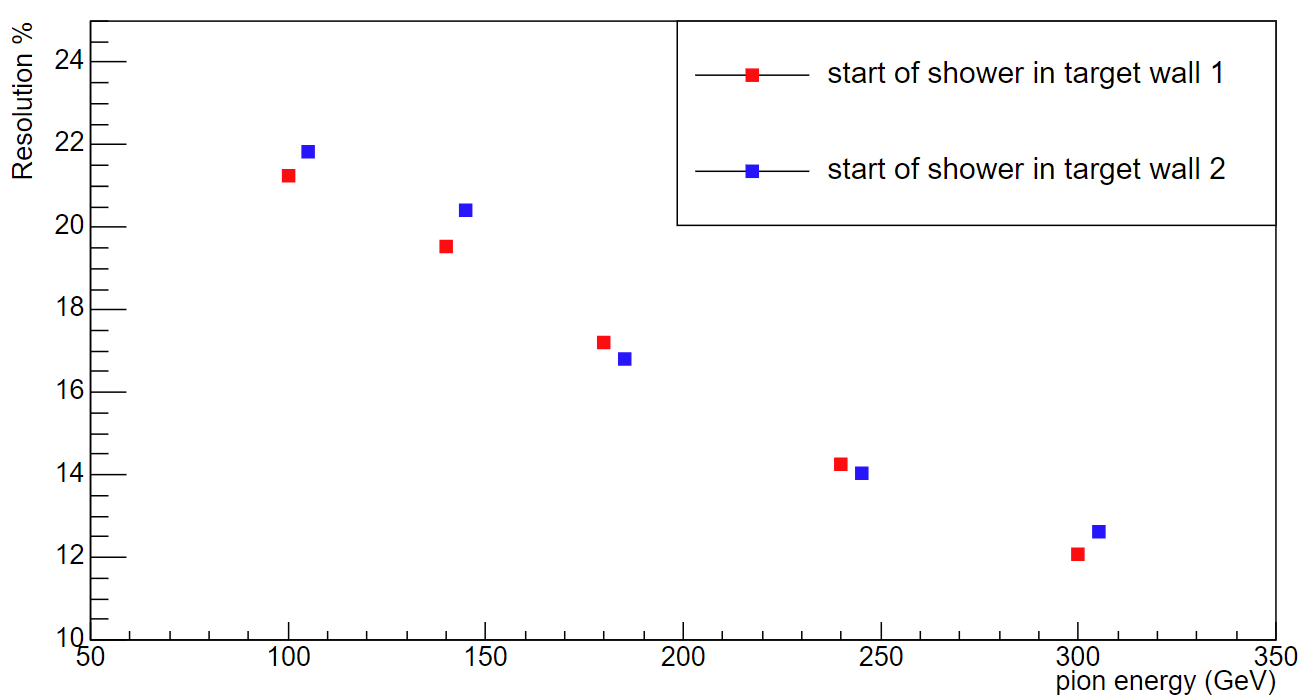}
\caption{Top: Relative offset between reconstructed energy $Etot$ and pion beam energy $Ebeam$. The error bars are the standard deviation of the reconstructed energy divided by the beam energy. Bottom: Energy resolution (standard deviation of the reconstructed energy divided by the beam energy) as a function of the pion beam energy.
}
\label{fig:Offset_vs_E}
\end{figure}

The data of the test samples are then reconstructed with these same constants. The standard deviation of the Gaussian fit to the core of each reconstructed energy distribution gives an estimate of the energy resolution (figure~\ref{fig:Ereco_st3}). 
One observes that for energies larger than 140~GeV the distributions have a tail from showers reconstructed with lower energies. This is due to showers with large energy deposits in the first US station causing  
SiPM saturation. This contribution grows with the shower energy: it is below $1\%$ at 140~GeV and reaches about $10\%$ at 300~GeV in the test beam detector which has a reduced target depth.
The effect is studied in details
in section~\ref{subsec:LateShowers}.
The relative offset between the reconstructed energy and the nominal beam energy for each data sample is summarized in figure~\ref{fig:Offset_vs_E}, where the error bars represent the standard deviation of the reconstructed energy distribution divided by the nominal energy. The reconstructed energy for showers starting in SciFi2 and SciFi3 is consistent with the true hadron energy within $\pm~10~\%$ across the whole energy range. 
The achieved energy resolution is shown in figure~\ref{fig:Offset_vs_E},
ranging from $22\%$ at $100$~GeV to $12\%$ at $300$~GeV.

\subsection{Energy Measurement in Late Showers}
\label{subsec:LateShowers}

Late showers originate near the end of the target depth; they are sampled by only one SciFi station placed downstream of the target, in both X and Y projections. 

The PCA determination of the calibration constants is prone to large uncertainties. Two issues affect the data:
(i) the single SciFi measurement (SciFi4 in the test beam setup) has substantial shower-to-shower fluctuations;
(ii) the US measurement is skewed due to SiPMs saturation.

About the $k$ calibration constant of SciFi,  it should be noted that the 
SciFi4 data are used also in ordinary showers, together with the SciFi2 and SciFi3 measurements. The conversion of QDC counts to energy has to be the same in either situation. $k$ must be the same as for ordinary showers.

\begin{figure}[tb]
\centering
\includegraphics[width=0.65\textwidth]{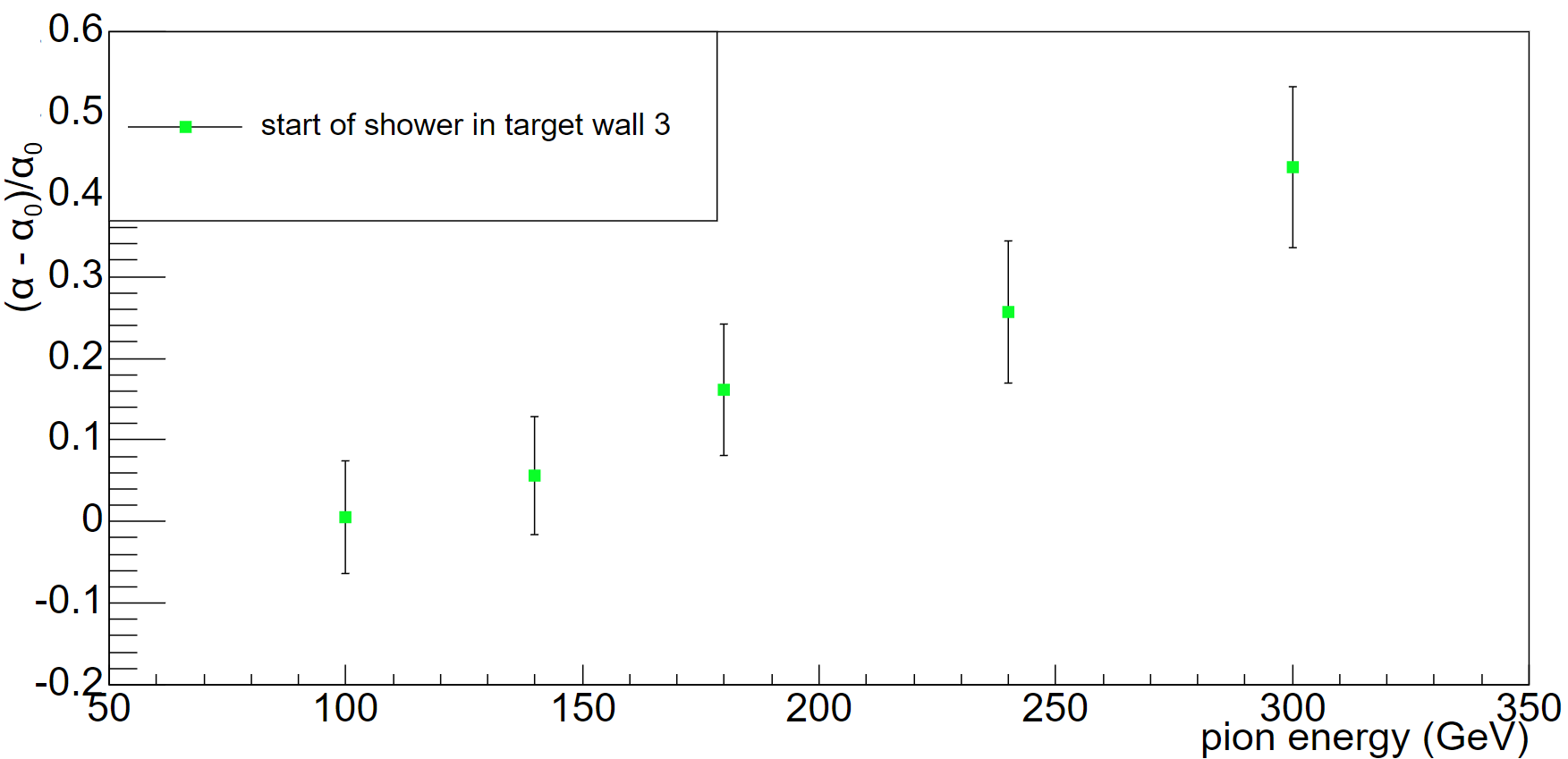}
\caption{  Relative variation of US $\alpha$ for late showers, originated in target wall 3 (tagged by only SciFi4), with respect to the $\alpha_0$ for ordinary showers, as function of the pion beam energy, 
while keeping the SciFi calibration parameter $k$ constant.}
\label{fig:alpha_saturation_scifi4}
\end{figure}

For the US system, the $\alpha$ calibration constant is significanlty modified since SiPMs saturation sets a hard limit to the QDC response. 
Figure~\ref{fig:alpha_saturation_scifi4}
shows how the best value of US $\alpha$, when fitting the data while keeping constant SciFi $k$, is modified   with respect to the $\alpha$ for ordinary showers 
($\alpha_0$) as function of the hadron energy. 
The deficit in $QDC_{US}$ can become as large as $40\%$ for a 300 GeV pion.

\begin{figure}[tb]
\centering
\includegraphics[width=0.6\textwidth]{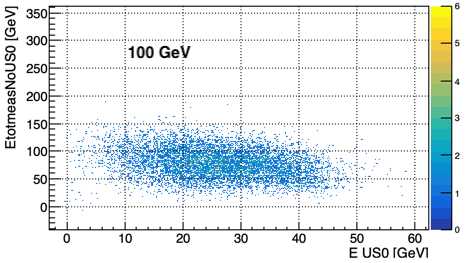}
\includegraphics[width=0.6\textwidth]{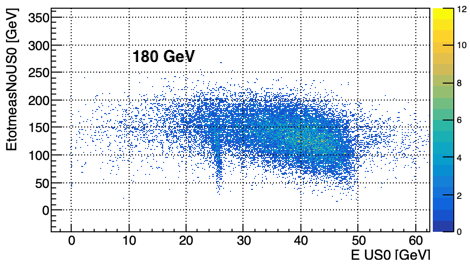}
\includegraphics[width=0.6\textwidth]{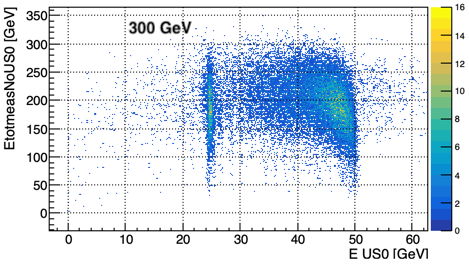}
\caption{ Scatter plots of measured shower energies excluding the contribution from US0 ( the first HCAL station downstream the target) versus the energy measured in US0, for 100, 180 and 300 GeV pions interacting 
in the final section of the target (wall 3 in the test beam setup).}
\label{fig:EtotnoUS0_vs_EUS0}
\end{figure}

\begin{figure}[tb]
\centering
\includegraphics[width=0.65\textwidth]{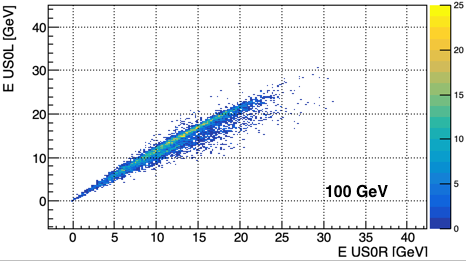}
\includegraphics[width=0.65\textwidth]{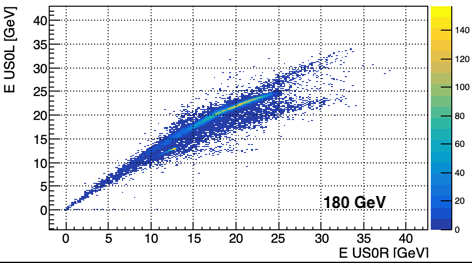}
\includegraphics[width=0.65\textwidth]{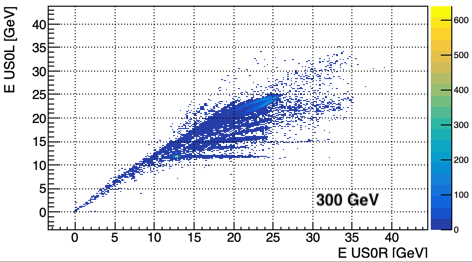}
\caption{ Scatter plots of measured shower energies at the left and right end of scintillating bars in US0 ( the first HCAL station downstream the target), for 100, 180 and 300 GeV pions interacting 
in the final section of the target (wall 3 in the test beam setup).}
\label{fig:EUS0L_vs_EUS0R}
\end{figure}

Of course,
a correction to $\alpha$ depending on the hadron energy is not applicable to SND@LHC data, in which the shower energy itself is the unknown quantity one aims to measure. 
However, it is reasonable to expect that SIPMs saturation will be maximal in US0, the first US station lying about $1~\lambda_{\text{INT}}$ downstream the target. 
At this stage, 
the shower transverse dimension is still limited, mostly contained in one or two scintillating bars, so that only a few SIPMs are fired by large energy deposits and saturated. When reaching US1, $1~\lambda_{\text{INT}}$ more downstream, the shower is more open and the energy shared over more bars and SIPMs, and saturation less probable. Then, using the linearity in the relation 
\begin{equation}
\begin{aligned}
\label{eq:Eline2}
Etot &= k \times QDC_{SciFi} + \alpha \times QDC_{US} \\
&= k \times QDC_{SciFi} + \alpha \times QDC_{US0}+ \alpha \times QDC_{US1+2+3+4}
\\
\end{aligned}
\end{equation} 
one could envisage to calculate a correction to $\alpha$ of US0, for the would-be signal above saturation, by using  all that is measured in the rest of the detector, except US0, as a rough shower energy estimator. 
This is investigated in figure~\ref{fig:EtotnoUS0_vs_EUS0} and
~\ref{fig:EUS0L_vs_EUS0R}:
it is observed that the energy measured in US0 grows until it reaches a limit beyond which all involved SiPMs are saturated. 
There are significant differences among SiPMs, some having lower saturation thresholds than others. 
Hence,
a correction to US0 $\alpha$ for the test beam detector is not directly applicable for the US0 SiPMs of the SND@LHC detector, and it should be calculated ad-hoc. 

It should be stressed that this procedure would correct the average energy response; shower-by-shower it would merely provide a very approximate energy measurement, with a large uncertainty. For showers originating late in the target depth, only a lower limit of the shower energy can be reliably established.


\section{Tuning of the Monte Carlo Simulation}
\label{sec:MCTuning}

The Monte-Carlo simulation of HCAL is based on the \texttt{sndsw} offline software package, which was initially developed by the SHiP~\cite{SHiP} collaboration. It utilizes the FairRoot~\cite{AlTurany2012} software for modelling the detector geometry. \texttt{sndsw} combines several high-energy physics software packages that allow for realistic simulation of both signal and background. These packages include \texttt{FLUKA}~\cite{Bhlen2014,Ahdida2022} and \texttt{GENIE}~\cite{ANDREOPOULOS201087} for neutrino generation and interaction, \texttt{PYTHIA6}~\cite{Sjstrand2006}, \texttt{DPMJET3}~\cite{Roesler2001} (Dual Parton Model, including charm) and \texttt{PYTHIA8}~\cite{Sjstrand2015} for the background from muon deep-inelastic scattering (DIS), and \texttt{GEANT4}~\cite{Agostinelli2003} for the particle propagation through the detector material. 
The main goal of modeling the test beam experiment is 
to tune the detector response in simulation to the experimental data, which will allow us to interpolate the energy calibration between the measured energies and extrapolate beyond.

\subsection{Simulation Description}
\label{subsec:Simulation}

\begin{figure}[tb]
    \centering
    \includegraphics[width=0.35\textwidth]{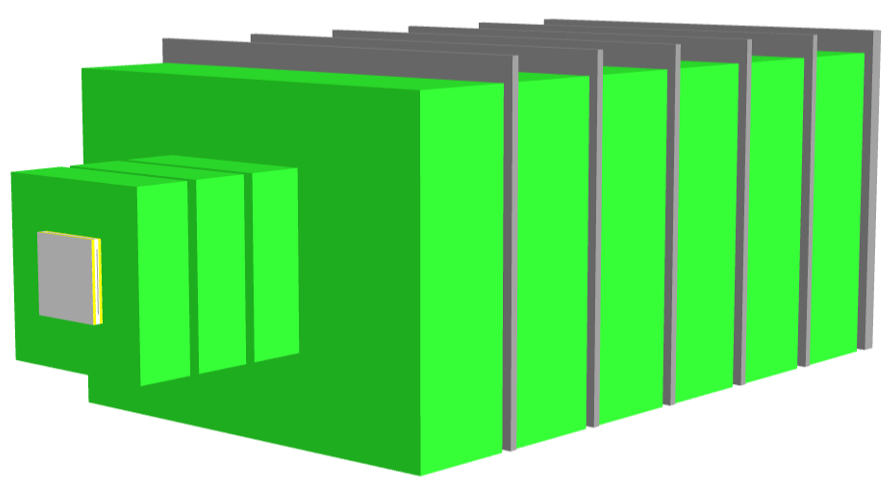}
    \caption{\texttt{sndsw} model of the test beam detector.}
    \label{testbeam_exp}
\end{figure}

 The geometry used to describe the detector setup in simulation employed the same materials that were used in the test beam experiment (see figure~\ref{testbeam_exp}). 
We simulated four datasets with $10^5$ events each of mono energetic pion beams with $Ebeam$ 100, 180, 300 and 500 GeV.
 The beam aimed at the center of the SciFi acceptance as in the test beam.
 All the information about the kinematics of particles in the detector (including energy losses, hit coordinates and timestamps) is stored in a \texttt{ROOT} output file.

Digitization is done after the main step of the simulation of the particle propagation through the detector. The procedure comprises summing energy losses of all particles within the individual sensitive volumes.
It is then assumed that the amount of produced scintillation photons is proportional to the energy loss in the corresponding scintillating bar and that the number of photons reaching each bar end is reduced by the photon attenuation length, as estimated from former test beam measurements~\cite{Acampora2024}.
In addition, it is assumed that the light detected in each SiPM is on average the light reaching the bar end divided by the 
number of large SiPMs ($N_{\text{SiPMs}}$) at the bar end.
The formula that describes the signal of the individual SiPM in the upstream part of HCAL is the following:
\begin{equation}
\begin{aligned}
    \text{Signal}_{\text{SiPM}} = \frac{\sum_{\text{particle}}{E_{\text{particle}}^{\text{loss}}}\times e^{-\Lambda \cdot d_{\text{particle}}}}{2\cdot N_{\text{SiPMs}}}.
\\
\end{aligned}
\end{equation}
Here $d_{\text{particle}}$ is the distance to the bar end depending on where (left or right end of a bar) the SiPM is located. $\Lambda$ is the light attenuation length calculated using the data of the previous test beam experiment~\cite{Acampora2024}. $\Lambda$ was estimated in a test beam by measuring the average signal produced in the electronics as a function of the position of the hit along the detector bars.

\subsection{Shower Tagging}

\begin{figure}[tb]
    \centering
    \includegraphics[width=.65\textwidth]{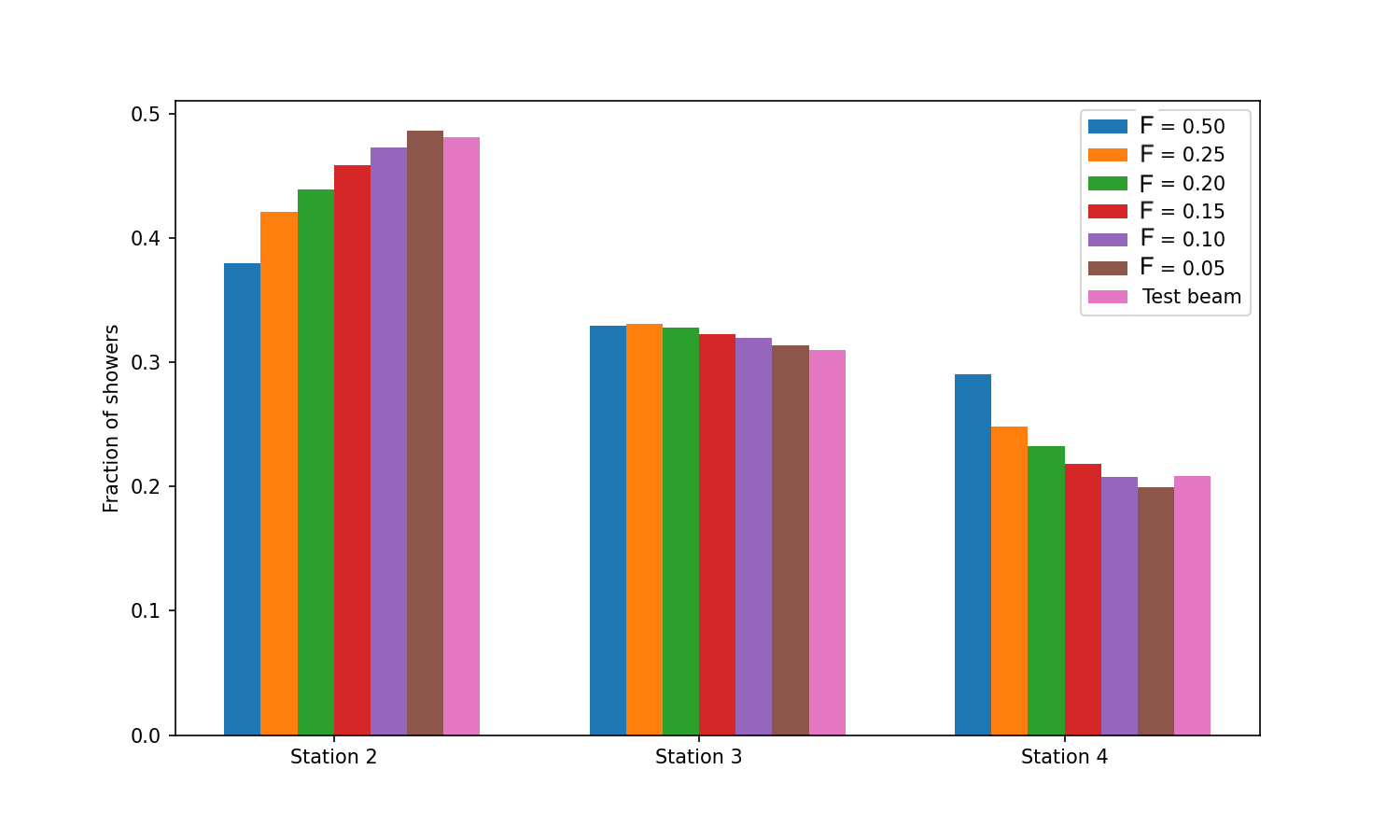}
    \caption{Shower origin tagging in the MC simulation. The fraction of tagged showers in each wall relative to all tagged showers}
    \label{tagging}
\end{figure}

A procedure was implemented to define the start of the hadronic shower in the simulation ("shower origin tagging") similarly to the data. We consider that the shower starts in wall $i$ if 
\begin{equation}
\begin{aligned}
        \frac{|E_{\text{upstream}}^{i} - E_{\text{downstream}}^{i}|}{E_{\text{upstream}}^{i}} > F
\\
\end{aligned}
\end{equation}
where $E_{\text{upstream}}^{i} \text{ and } E_{\text{downstream}}^{i}$ are the sums of particle energies in the SciFi planes upstream and downstream of the wall~$i$. $F$ is an adjustable parameter that can be fixed according to the shower tagging in the experiment (see section~\ref{subsec:ShowerTagging}). We considered several $F$ values in a range between $0.05$ and $0.5$ to find the distribution of 300 GeV shower origins closest to the experimental one comparing with the probability theory prediction. As can be seen from
figure~\ref{tagging}, the value of $F \sim 0.10$ provides the results that best correspond to the experimental data. Hence, we set $F = 0.10$.

\subsection{Energy Calibration}
\label{subsec:MCEnergyCalibration}

The energy calibration procedure for Monte-Carlo simulated data is similar to the one for the experimental data. However, instead of using the total signal per event registered in the SciFi and US detectors, in the calibration we used the total energy losses per event in these detectors to obtain  energy response and resolution that represent an ideal case. 

\begin{figure}[tb]
    \centering
    \includegraphics[width=1.0\textwidth]{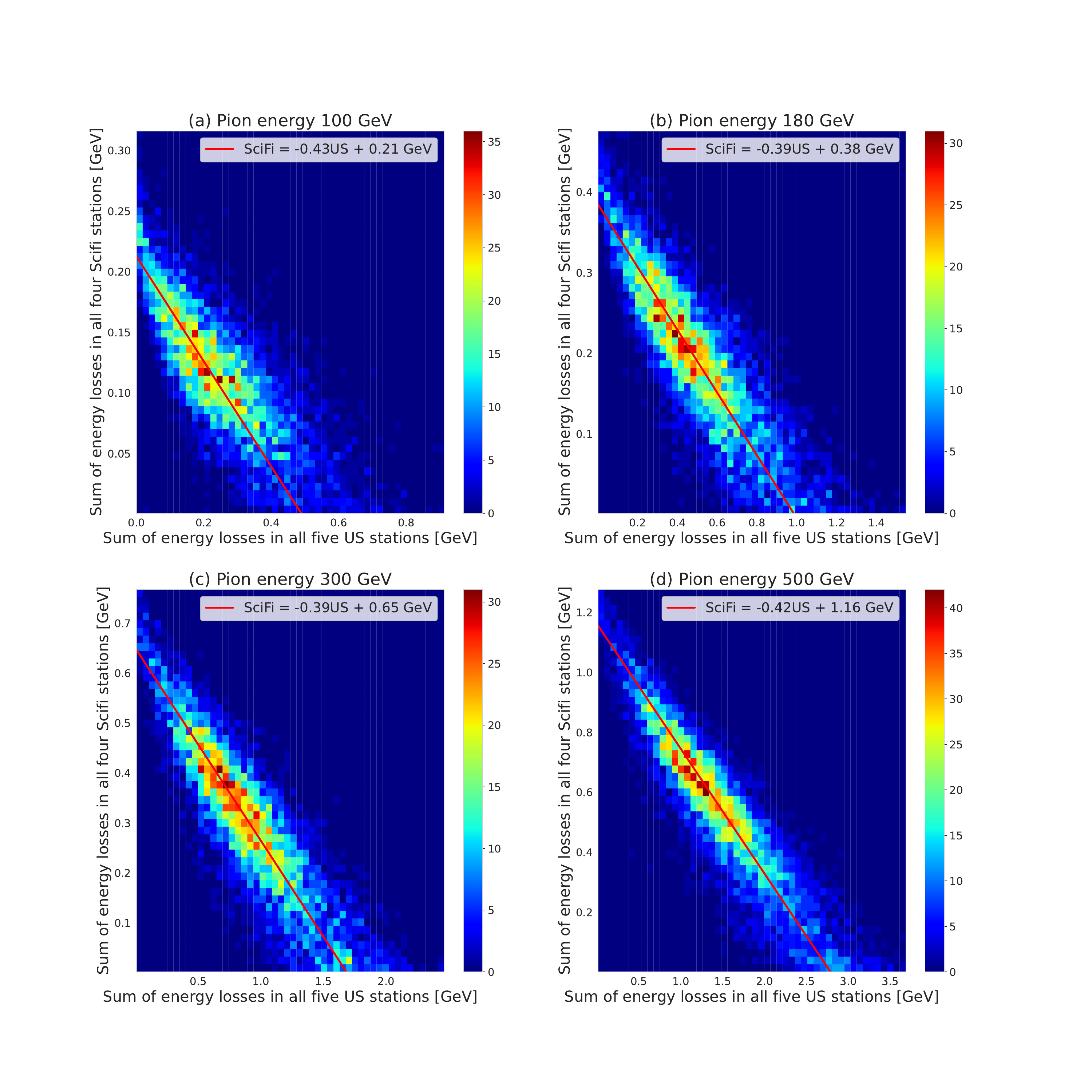}
    \caption{
     In the MC simulation, energy losses in SciFi and US stations, for pion beams of three benchmark energies $Ebeam$ (a) 100 GeV, (b) 180 GeV and (c) 300 GeV, and for (d) 500 GeV with energy calibration best-fit line superimposed.
}
    \label{calib1}
\end{figure}

The energy loss distributions for three benchmark energies (100, 180 and 300 GeV) and for 500 GeV are shown in figure~\ref{calib1}.
Given the inverse linear dependency between total energy losses in SciFi and US, we fitted this 2D SciFi-US distribution by a linear function $\text{SciFi} = A\cdot\text{US} + B$, where $A$ and $B$ are fitting parameters and \text{SciFi} and \text{US} are total energy losses, for various initial pion energies. 
A and B are equivalent to $-\alpha/k$ 
and to $Etot/k$, respectively, in the test beam data analysis, 
where $Etot$ is the reconstructed pion energy
(see section~\ref{subsec:HadronicEnergyMeasurement}).

\begin{figure}[tb]
    \centering
   \includegraphics[width=0.9\textwidth]{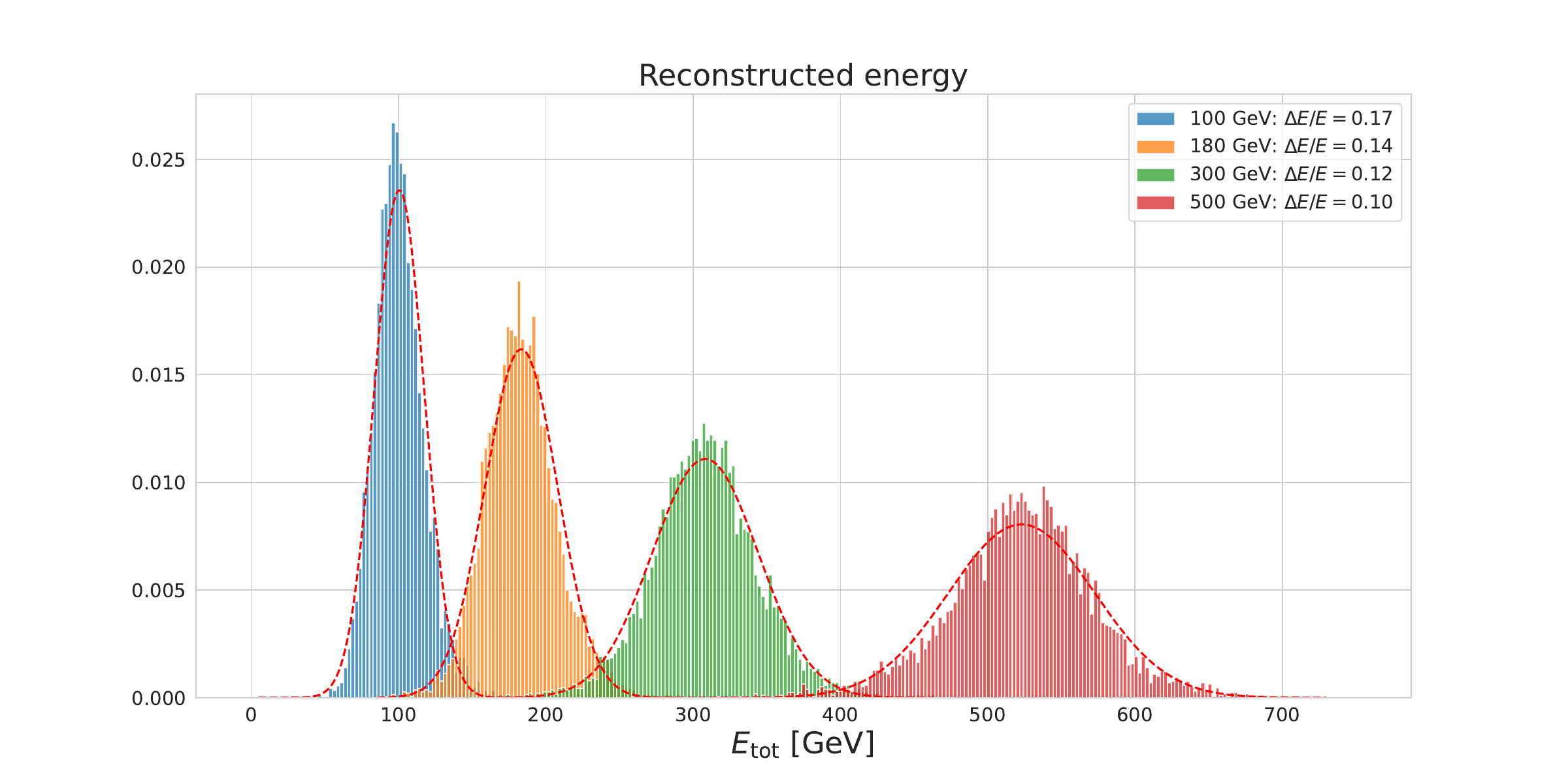}
    \caption{ MC simulation distributions in the reconstructed total energy $Etot$ for pion energies of 100, 180, 300 and 500 GeV and the SND@LHC test beam setup.
}
    \label{calib2}
\end{figure}

\begin{figure}[tb]
    \centering
    \includegraphics[width=1.0\textwidth]{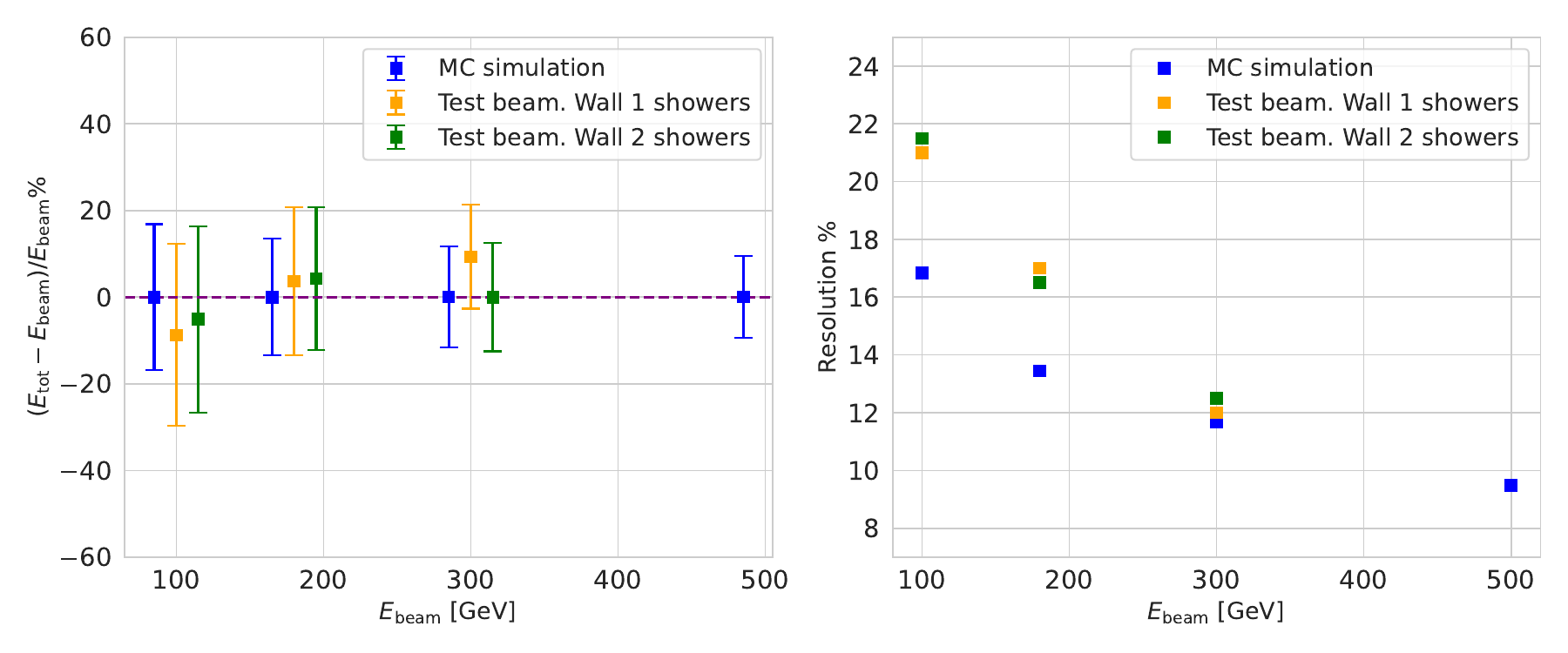}
    \caption{On the left, relative offset between reconstructed energy $Etot$ and initial pion energy $Ebeam$, and, on the right, energy resolution, in the ideal case of simulation, compared to the results of the test beam data analysis for "ordinary" showers originated in target walls 1 and 2.}
    \label{reco}
\end{figure}

We observe, as expected,  that $A$ does not change significantly with increasing pion beam energy $Ebeam$ and that B grows proportionally to $Ebeam$.
The reconstructed pion energy $Etot$ distributions are also shown in
figure~\ref{calib2}.
In contrast with figure~\ref{fig:Ereco_st3}, the distributions have symmetric tails, since there is no SiPM saturation in the simulation.  
The energy resolution values are given by the standard deviations of the Gaussian fits on the energy distributions and are 17$\%$, 15$\%$ and 13$\%$ respectively (see figure~\ref{reco}). 

The energy calibration performed in the test beam experiment resulted, for "ordinary" showers, in a energy resolution in reasonable agreement with the Monte-Carlo ideal case.
A discrepancy is observed at lower energies as expected, since, in the simulation, electronic noise is not present and signal thresholds are kept unrealistically low.
Also note that the reconstructed energies for showers originating late in the target are not biased in the Monte-Carlo simulation as observed in the experimental data because of SiPMs saturation.


\section{Summary}
\label{sec:conclusion}

The SND@LHC experiment detects high energy neutrinos emerging from LHC collisions at very small angles.
The detector consists of two sections: a target instrumented with SciFi stations, and a hadronic calorimeter HCAL equipped with scintillating bars.
Neutrino interactions in the target produce hadronic showers.
In order to calibrate the energy measurement, a replica of the SND@LHC apparatus was exposed to hadron beams of 100,\,140,\,180,\,240,\,and~300~GeV at the CERN SPS H8 line.
The HCAL was an exact replica of the SND@LHC hadron calorimeter. The target section consisted of three of the five walls of the SND@LHC target.

The experimental setup allowed for studying the dependence of the energy sharing between target and calorimeter upon the position of shower origin along the target depth.
Showers are tagged using the density of in-time hits in the SciFi planes, and thus the origins of individual showers can be located in the target sections interleaving the  SciFi stations.
Showers are then categorized by their origins in the target; in each category the shower total energy is obtained from a linear sum of the energy releases in the target and in the calorimeter, estimated from the signal generated in the SciFi stations and in the HCAL scintillating bars.
By using a Principal Component Analysis (PCA) fit, conversion factors from measured values of the digitized charge (QDC) to energy in GeV are calculated for both SciFi and HCAL. Hadronic showers are then grouped in two topologies, named “ordinary” and “late”.

Ordinary showers are sampled in the target region by at least two SciFi stations in both X and Y projections. The energy resolution is determined to be about $20~\%$ at 100~GeV and $12~\%$ at 300~GeV. 

This behavior is in agreement with the expectations of the Monte Carlo simulation.
The calibration constants obtained with the test beam setup apply to showers recorded by the SND@LHC experiment when the shower origins are in Wall~1 to Wall~4 of the five sections in which the target is subdivided. 
However, rescaling factors
for SciFi 
are to be calculated due to the smaller lateral shower radius in tungsten and to the larger size - i.e. longer fibers - of the SciFi modules. 
Systematic uncertainties in the extrapolation to energies smaller than 100~GeV should be carefully evaluated.

Late showers originate near the end of the target depth and are sampled by only the one SciFi station downstream the target in both X and Y projections. 
The PCA fit has large uncertainties, due to the 
significant shower-to-shower dispersion of the single SciFi measurement and to the saturation of the SiPMs of the scintillating bars in the first station of HCAL for narrow energetic showers. 
Only a lower limit of the shower energy can be reliably established for showers originating late in the target depth.  
Concerning the SND@LHC experiment, this result implies that,
although showers can be efficiently tagged in all five Walls of the target, only interactions in target Wall 1 to 4 should be considered
for measuring the energy distribution of hadronic showers in neutrino events.


\acknowledgments

We are grateful to M. Van Dijk and A. Baratto Roldan for optimizing the SPS H8 beam line for our goals and for their continuous assistance
during the data taking, E. B. Holzer, M. R. Jaekel and D. Lazic for their help with scheduling the run and to S. Ozkorucuklu for help during the preparation of the experiment.

We acknowledge the support  provided by the following funding agencies:  CERN;  the Bulgarian Ministry of Education and Science within the National
Roadmap for Research Infrastructures 2020–2027 (object CERN); ANID FONDECYT grants No. 3230806, No. 1240066, 1240216 and ANID  - Millenium Science Initiative Program -  $\rm{ICN}2019\_044$ (Chile); the Deutsche Forschungsgemeinschaft (DFG, ID 496466340); the Italian National Institute for Nuclear Physics (INFN); JSPS, MEXT, the~Global COE program of Nagoya University, the~Promotion
and Mutual Aid Corporation for Private Schools of Japan for Japan;
the National Research Foundation of Korea with grant numbers 2021R1A2C2011003, 2020R1A2C1099546, 2021R1F1A1061717, and
2022R1A2C100505; Fundação para a Ciência e a Tecnologia, FCT (Portugal), 
CERN/FIS-INS/0028/2021; the Swiss National Science Foundation (SNSF); TENMAK for Turkey (Grant No. 2022TENMAK(CERN) A5.H3.F2-1).
M.~Climesu, H.~Lacker and R.~Wanke are funded by the Deutsche Forschungsgemeinschaft (DFG, German Research Foundation), Project 496466340. We acknowledge the funding of individuals by Fundação para a Ciência e a Tecnologia, FCT (Portugal) with grant numbers  CEECIND/01334/2018, 
CEECINST/00032/2021 and 
PRT/BD/153351/2021. 
This work has been partially supported by Spoke 1 "FutureHPC \& BigData" of ICSC - Centro Nazionale di Ricerca in High-Performance-Computing, Big Data and Quantum Computing, funded by European Union - NextGenerationEU.





\end{document}

%% file: Authorlist.tex
\author[9]{D.~Abbaneo~\orcidlink{0000-0001-9416-1742}}
\author[42]{S.~Ahmad~\orcidlink{0000-0001-8236-6134}}
\author[1,2]{R.~Albanese~\orcidlink{0000-0003-4586-8068}}
\author[1]{A.~Alexandrov~\orcidlink{0000-0002-1813-1485}}
\author[1,2]{F.~Alicante~\orcidlink{0009-0003-3240-830X}}
\author[1,2]{F.~Aloschi~\orcidlink{0000-0002-2501-7525}}
\author[6]{K.~Androsov~\orcidlink{0000-0003-2694-6542}}
\author[3]{A.~Anokhina~\orcidlink{0000-0002-4654-4535}}
\author[38]{C.~Asawatangtrakuldee~\orcidlink{0000-0003-2234-7219}}
\author[32,27]{M.A.~Ayala~Torres~\orcidlink{0000-0002-4296-9464}}
% ...

\affiliation[1]{Sezione INFN di Napoli, Napoli, 80126, Italy}
\affiliation[2]{Universit\`{a} di Napoli ``Federico II'', Napoli, 80126, Italy}
\affiliation[3]{Affiliated with an institute formerly covered by a cooperation agreement with CERN}
% ...
\affiliation[6]{Institute of Physics, EPFL, Lausanne, 1015, Switzerland}
% ...
\affiliation[9]{European Organization for Nuclear Research (CERN), Geneva, 1211, Switzerland}
% ...
\affiliation[27]{Millennium Institute for Subatomic physics at high energy frontier-SAPHIR, Santiago, 7591538, Chile}
% ...
\affiliation[32]{Center for Theoretical and Experimental Particle Physics, Facultad de Ciencias Exactas, Universidad Andr\`es Bello, Fernandez Concha 700, Santiago, Chile}
% ...
\affiliation[38]{Chulalongkorn University, Bangkok, 10330, Thailand}
% ...
\affiliation[42]{Affiliated with Pakistan Institute of Nuclear Science and Technology (PINSTECH), Nilore, 45650, Islamabad, Pakistan}
% ...